\RequirePackage{fix-cm}
\RequirePackage{fixltx2e}
\documentclass[french,english]{article}
\usepackage[T1]{fontenc}
\usepackage[latin9]{inputenc}
\usepackage{geometry}
\usepackage[active]{srcltx}
\usepackage{color}
\usepackage{babel}
\makeatletter
\addto\extrasfrench{%
   \providecommand{\fg}{\ifdim\lastskip>\z@\unskip\fi~\frqq}%
}

\makeatother
\usepackage{array}
\usepackage{float}
\usepackage{multirow}
\usepackage{amsmath}
\usepackage{amsthm}
\usepackage{amssymb}
\usepackage{graphicx}
\usepackage{setspace}
\usepackage[authoryear]{natbib}
\usepackage[unicode=true,pdfusetitle,
 bookmarks=true,bookmarksnumbered=true,bookmarksopen=true,bookmarksopenlevel=1,
 breaklinks=true,pdfborder={0 0 1},backref=false,colorlinks=true]
 {hyperref}
\hypersetup{
 linkcolor=blue, citecolor=blue, urlcolor=blue, filecolor=blue,pdfpagelayout=OneColumn, pdfnewwindow=true,pdfstartview=XYZ, plainpages=false, pdfpagelabels, hyperindex=true}
\usepackage{breakurl}

\makeatletter

\providecommand{\tabularnewline}{\\}
\newcommand{\lyxdot}{.}

\floatstyle{ruled}
\newfloat{algorithm}{tbp}{loa}
\providecommand{\algorithmname}{Algorithm}
\floatname{algorithm}{\protect\algorithmname}

\numberwithin{equation}{section}
\numberwithin{figure}{section}
\numberwithin{table}{section}

\usepackage{pdflscape}
\usepackage{fancyhdr}
\usepackage{lastpage}
\usepackage{amsmath}
\usepackage{graphicx}
\usepackage{color}
\usepackage{fancybox} 
\usepackage{moreverb} 
\usepackage{listings} 
\usepackage{backref}
\usepackage{algorithmic}
\usepackage{enumitem}

\usepackage{ifpdf} 
\ifpdf 

 \IfFileExists{lmodern.sty}{\usepackage{lmodern}}{}

\fi 

\let\myTOC\tableofcontents
\renewcommand\tableofcontents{%
  \pdfbookmark[1]{\contentsname}{}
  \myTOC
}

\def\LyX{\texorpdfstring{%
  L\kern-.1667em\lower.25em\hbox{Y}\kern-.125emX\@}
  {LyX}}

\@addtoreset{footnote}{section}

\renewcommand*{\backref}[1]{}
\renewcommand*{\backrefalt}[4]{%
   \ifcase #1 
    \or 
      (Cited on page~#2)%
   \else
      (Cited on pages~#2)
    \fi} 

\usepackage{dsfont}
\usepackage{bbm}

\usepackage{eurosym}

\usepackage[algo2e,ruled,vlined]{algorithm2e}
\usepackage{xcolor}
\definecolor{algoColorKeyword}{named}{blue}
\definecolor{com}{rgb}{0,0.6,0.3}
\definecolor{algoColorComment}{named}{com}

\SetKwComment{Comment}{\!\%\,}{}
\usepackage{setspace}
\usepackage{chngcntr}
\counterwithout{figure}{section}
\counterwithout{table}{section}

\makeatother

\addto\captionsfrench{\renewcommand{\algorithmname}{Algorithme}}

\begin{document}

\title{Dynamic Portfolio Optimization with Liquidity Cost and Market Impact:
A Simulation-and-Regression Approach}

\author{Rongju Zhang\thanks{Corresponding author. Email: \protect\href{mailto:rongju.zhang@monash.edu}{rongju.zhang@monash.edu}.
School of Mathematical Sciences, Monash University}, Nicolas Langren\'e\thanks{RiskLab, CSIRO}, Yu Tian\thanks{School of Mathematical Sciences, Monash University},
Zili Zhu\thanks{RiskLab, CSIRO}, Fima Klebaner\thanks{School of Mathematical Sciences, Monash University}
and Kais Hamza\thanks{School of Mathematical Sciences, Monash University}}

\date{{\normalsize{}First version: October 25, 2016}\\
{\normalsize{}This revised version: August 28, 2017}}
\maketitle
\begin{abstract}
We present a simulation-and-regression method for solving dynamic
portfolio allocation problems in the presence of general transaction
costs, liquidity costs and market impacts. This method extends the
classical least squares Monte Carlo algorithm to incorporate switching
costs, corresponding to transaction costs and transient liquidity
costs, as well as multiple endogenous state variables, namely the
portfolio value and the asset prices subject to permanent market impacts.
To do so, we improve the accuracy of the control randomization approach
in the case of discrete controls, and propose a global iteration procedure
to further improve the allocation estimates. We validate our numerical
method by solving a realistic cash-and-stock portfolio with a power-law
liquidity model. We quantify the certainty equivalent losses associated
with ignoring liquidity effects, and illustrate how our dynamic allocation
protects the investor's capital under illiquid market conditions.
Lastly, we analyze, under different liquidity conditions, the sensitivities
of certainty equivalent returns and optimal allocations with respect
to trading volume, stock price volatility, initial investment amount,
risk-aversion level and investment horizon.

\vspace{1mm}
$\,$

\vspace{1mm}

$\,$

\vspace{1mm}

$\,$
\end{abstract}
\newpage{}

\section{Introduction}

The effect of liquidity on the design of dynamic multi-period portfolio
selection methods (a.k.a. asset allocation, portfolio optimization
or portfolio management) has drawn great attention from academics
and practitioners alike. Liquidity affects portfolio allocation in
two main ways: temporary liquidity cost and permanent market impact.
Liquidity cost, also known as implementation shortfall, temporary
market impact or transitory market impact, is the difference between
the realized transaction price and the pre-transaction price. Market
impact is the permanent shift in the asset price after a transaction,
due to the post-transaction ``resilience'' of the limit order book.
These liquidity effects depend on several factors, such as the nature
of the exchange platform, the duration of the trade execution, the
transaction volume, the asset volatility and so on. Up to now, liquidity
modeling for dynamic portfolio selection has been impeded by the intractability
of analytical solutions and by the limited capability of numerical
methods to handle endogenous stochastic prices. The purpose of the
present paper is to introduce a new simulation-and-regression method
capable of handling multivariate portfolio allocation problems under
general transaction costs, liquidity costs and market impacts. 

The original literature on dynamic portfolio selection started with
simple problems without transaction costs. The seminal papers, \citet{Mossin1968},
\citet{Samuelson1969}, \citet{Merton1969} and \citet{Merton1971}
provide closed-form solutions of optimal asset allocation strategies
for long-term investors. In reality though, every transaction incurs
commission fee (or brokerage cost), and several improvements have
therefore been proposed to account for transaction cost. Examples
of closed-form solutions are \citet{Davis1990}, \citet{Shreve1994},
\citet{Liu2004} and \citet{Garleanu2013}. Examples of numerical
methods are \citet{Lynch2010}, \citet{Muthuraman2008} and \citet{Brown2011}.
Transient liquidity cost, viewed as another type of transaction cost,
has also been studied by many researchers in the context of dynamic
portfolio selection problems. \citet{Cetin2007} show the existence
of optimal portfolios and how to turn the marginal price process under
the optimal strategy into a martingale using the optimal terminal
wealth as change of measure. We refer to \citet*{Ma2013} and \citet{Lim2014}
for examples of solving the Hamilton\textendash Jacobi\textendash Bellman
(HJB) equation. Other than liquidity cost, permanent market impact
is also a crucial element when dealing with large transactions, as
it affects portfolio valuation due to the shifts in asset prices.
This effect has been widely incorporated in the studies of portfolio
liquidation problems. For example, \citet{Bertsimas1998}, \citet{Almgren2000},
\citet{Obizhaeva2013} and \citet*{Tsoukalas2015}. These works, although
restricted to either linear or linear-quadratic objective functions,
provide a broad overview of trading modeling in illiquid markets.
Dynamic portfolio selection under permanent market impact has been
formulated in \citet*{Vath2007} as an impulse control problem under
state constraints, where the authors characterize the value function
as the unique constrained viscosity solution to the associated quasi-variational
HJB inequality. This framework has been extended to numerical approximation
in \citet*{Gaigi2016}. \citet{Garleanu2013} derive a closed-form
optimal portfolio policy for the mean-variance framework with quadratic
transaction costs such that liquidity cost and market impact are included.
Following on this framework, many extensions have been proposed, for
example \citet*{Collin-Dufresne2015} and \citet*{Mei2016}. However,
due to the analytically intractable formulation, these methods are
restricted in the range of applications when market impacts are present.

To broaden the range of applications, the least-squares Monte Carlo
(LSMC) algorithm is a possible solution. The LSMC algorithm, originally
developed by \citet{Carriere1996}, \citet{Longstaff2001} and \citet{Tsitsiklis01}
for the pricing of American options, has been extended to solve dynamic
portfolio selection problems in \citet*{Brandt2005}, \citet{Garlappi2010}
and \citet{Cong2016}. \citet{Brandt2005} determine a semi-closed
form by solving the first order condition of the Taylor series expansion
of the future value function. \citet{Garlappi2010} claim that the
convergence of \citet{Brandt2005}\textquoteright s method is not
stable and that it cannot handle problems where the control variable
depends on the endogenous wealth variable. Instead, they introduce
a state variable decomposition method to overcome this drawback. However,
this decomposition relies on a linear separation between the observable
component and stochastic deviation of returns, which cannot be applied
to general return distributions. \citet{Cong2016} use a multi-stage
strategy to perform forward simulation of control variables which
are iteratively updated in the backward recursive program, where the
admissible control sets are constructed as small neighborhoods of
the solutions to the multi-stage strategy. Later, \citet{Cong2017}
combine \citet{Jain2015}'s stochastic bundling technique with \citet{Brandt2005}'s
method. To sum up, these three papers have opened the way to the use
of the LSMC algorithm for solving dynamic portfolio selection problems,
but are at this stage still limited and constrained in their possible
formulations of transaction cost, liquidity cost and market impact. 

In this paper, we make three contributions to this literature. Our
first contribution is to propose a LSMC algorithm to solve dynamic
portfolio selection problems with no restriction in the formulations
of transaction cost, liquidity cost and market impact, and allowing
for multiple assets with general dynamics in a computationally tractable
way. Our method is the most general and versatile available in the
literature, and can be easily adapted to other applications involving
optimal multiple switching problems. 

Our second contribution is to improve the numerical performance of
\citet*{Kharroubi2014}\textquoteright s control randomization algorithm
in the case of discrete control. In \citet{Kharroubi2014}, the randomized
controls are part of the regression inputs, and the regression basis
is extended accordingly. However, an inadequate regression basis for
the control variable can slow down the convergence of this approach,
all the more so for highly nonlinear payoffs. Moreover, finding an
adequate basis for the controls can be problematic in practice. To
avoid this difficulty, we account for the control information by discretizing
the control space and performing one regression per control level.
This discrete control approach extends the optimal switching approach
(\citet{Boogert2008}, \citet*{Aid14}) to the case with endogenous
state variables. Finally, we iterate the whole algorithm by replacing
the initial randomized controls by the optimal control estimates from
the previous run. We show that these combined modifications improve
the portfolio allocation estimates. 

Our third contribution is to present an empirical study on how dynamic
portfolio allocations are affected by transient and permanent liquidity
effects. We apply our method to solve a realistic cash-and-stock portfolio
allocation problem, for which we adopt the power-law liquidity model
of \citet*{Almgren2005}. We measure the certainty equivalent losses
associated with ignoring liquidity issues, and illustrate the ability
of our dynamic allocation to protect the investor's capital in illiquid
markets. Finally, based on different liquidity scenarios, we analyze
the sensitivity of certainty equivalent returns and portfolio allocations
with respect to trading volumes, stock price volatility, initial investment
amount, risk-aversion level and investment horizon. 

The outline of the paper is as follows. Section \ref{sec:portfolio-optimisation}
formulates our dynamic portfolio selection problem with transaction
cost, liquidity cost and market impact. Section \ref{sec:LSMC} describes
the LSMC algorithm developed to solve this problem. Section \ref{sec:liquidity}
describes the parametric liquidity model we used. Section \ref{sec:Application}
describes our numerical experiments and Section \ref{sec:Conclusion}
concludes the paper. 

\section{Problem Description\label{sec:portfolio-optimisation}}

In this section, we provide the detailed mathematical description
of the portfolio allocation problem we aim to solve. Consider a dynamic
portfolio selection problem over a finite time horizon $T$. Suppose
there are $d$ risky assets available for investment. Denote $r^{f}$
as the risk-free rate. Let $\left\{ \mathbf{r}_{t}\right\} _{0\leq t\leq T}=\left\{ r_{t}^{i}\right\} _{0\leq t\leq T}^{1\leq i\leq d}$
and $\left\{ \mathbf{S}_{t}\right\} _{0\leq t\leq T}=\left\{ S_{t}^{i}\right\} _{0\leq t\leq T}^{1\leq i\leq d}$
respectively denote the asset returns and prices. Denote $\left\{ \mathbf{Z}_{t}\right\} _{0\leq t\leq T}$
as the vector of return predictors. This vector $\left\{ \mathbf{Z}_{t}\right\} _{0\leq t\leq T}$
is used to construct the dynamics of the assets. Let $\boldsymbol{\alpha}_{t}=\left(\alpha_{t}^{i}\right)_{1\leq i\leq d}$
be the portfolio allocation in each risky asset at time $t$; the
allocation in the risk-free asset is then given by $\alpha_{t}^{f}=1-\sum_{1\leq i\leq d}\alpha_{t}^{i}$.
In a similar manner, let $\mathbf{q}_{t}=\left(q_{t}^{i}\right)_{0\leq t\leq T}^{1\leq i\leq d}$
describes the number of units held in each risky asset and let $\left\{ q_{t}^{f}\right\} _{0\leq t\leq T}$
denote the amount allocated in the risk-free cash. Define $\Delta q_{t}^{i}:=q_{t}^{i}-q_{t^{-}}^{i}$
as the transaction volume for the $i^{\text{th}}$ risky asset at
time $t$. Let $\mathcal{A}\subseteq\mathbb{R}^{d}$ be the set of
admissible portfolio strategies. These sets may include constraints
defined by the investor, such as weight limits in each individual
asset for example. Finally, let $\left\{ W_{t}\right\} _{0\leq t\leq T}$
denote the portfolio value (or wealth) process.

For every transaction, due to transaction cost, liquidity cost and
market impact, there are immediate shifts to the endogenous asset
prices and portfolio value. Let $\mathbf{TC}\left(\Delta\mathbf{q}_{t}\right)=\left\{ \textrm{TC}^{i}\left(\Delta q_{t}^{i}\right)\right\} _{1\leq i\leq d}$,
$\mathbf{LC}\left(\Delta\mathbf{q}_{t}\right)=\left\{ \textrm{LC}^{i}\left(\Delta q_{t}^{i}\right)\right\} _{1\leq i\leq d}$
and $\mathbf{MI}\left(\Delta\mathbf{q}_{t}\right)=\left\{ \textrm{MI}^{i}\left(\Delta q_{t}^{i}\right)\right\} _{1\leq i\leq d}$
respectively denote the vector of transaction costs, liquidity costs
and market impacts generated by the transaction $\Delta q_{t}^{i}$
for each risky asset $i=1,...,d$. In general, we write these quantities
as deterministic functions of transaction volume: $\textrm{TC}{}^{i}:\mathbb{R}\rightarrow\mathbb{R}$,
$\textrm{LC}{}^{i}:\mathbb{R}\rightarrow\mathbb{R}$ and $\textrm{MI}^{i}:\mathbb{R}\rightarrow\mathbb{R}$,
and thus $\textrm{\textbf{TC}}:\mathbb{R}^{d}\rightarrow\mathbb{R}^{d}$,
$\mathbf{LC}:\mathbb{R}^{d}\rightarrow\mathbb{R}^{d}$ and $\mathbf{MI}:\mathbb{R}^{d}\rightarrow\mathbb{R}^{d}$.
Given a transaction $\left\{ \Delta q_{t}^{i}\right\} _{1\leq i\leq d}$
at time $t$, the following immediate changes occur: 
\begin{eqnarray}
\mathbf{S}_{t} & = & \mathbf{S}_{t^{-}}+\mathbf{MI}\left(\Delta\mathbf{q}_{t}\right),\nonumber \\
W_{t} & = & W_{t^{-}}-\mathbf{TC}\left(\Delta\mathbf{q}_{t}\right)\cdot\vec{\mathbf{1}}_{d}-\mathbf{LC}\left(\Delta\mathbf{q}_{t}\right)\cdot\vec{\mathbf{1}}_{d}+\mathbf{MI}\left(\Delta\mathbf{q}_{t}\right)\cdot\mathbf{q}_{t}.\label{eq:wealth-switch}
\end{eqnarray}
where $\vec{\mathbf{1}}_{d}$ is a vector of size $d$ with all the
entries equal to $1$. 

It is important to note that there are two possible descriptions of
the portfolio positions:\emph{ absolute positions} using the quantity
(number of units) in each asset $\mathbf{q}_{t}$\emph{ }and\emph{
relative positions} using the proportions of wealth in each asset
$\boldsymbol{\alpha}_{t}$. We describe our portfolio allocation decisions
using $\boldsymbol{\alpha}_{t}$, while transaction cost, liquidity
cost and market impact depend on $\mathbf{q}_{t}$. Fortunately, there
is a natural one-to-one correspondence between these two descriptions,
namely, 
\begin{equation}
\boldsymbol{\alpha}_{t}\times W_{t}=\mathbf{q}_{t}\times\mathbf{S}_{t},\label{eq:q-alpha-relation}
\end{equation}
where ``$\times$'' denotes the element-wise multiplication and
we also denote ``$\div$'' as element-wise division. Suppose that
at time $t$, one wants to rebalance the portfolio from the absolute
position $\mathbf{q}_{t-1}$ to the relative weight $\boldsymbol{\alpha}_{t}\in\mathcal{A}$.
Then, using the dynamics \eqref{eq:wealth-switch} and the relation
\eqref{eq:q-alpha-relation}, the following system of equations holds:
\begin{align}
\boldsymbol{\alpha}_{t}\times\left(W_{t^{-}}-\mathbf{TC}(\Delta\mathbf{q}_{t})\cdot\vec{\mathbf{1}}_{d}-\mathbf{LC}(\Delta\mathbf{q}_{t})\cdot\vec{\mathbf{1}}_{d}+\text{\textrm{\textbf{MI}}}(\Delta\mathbf{q}_{t})\cdot\mathbf{q}_{t}\right) & =\mathbf{q}_{t}\times\left(\mathbf{S}_{t^{-}}+\textrm{\textbf{MI}}(\Delta\mathbf{q}_{t})\right).\label{eq:solve-q-alpha}
\end{align}
This is a system of nonlinear equations coupled by the wealth variable.
Solving these equations enables us to simultaneously update $\boldsymbol{\alpha}_{t}$
and $\mathbf{q}_{t}$, and thus avoid the potential mismatch between
actual allocation and target allocation. To solve it numerically (i.e.
being given $\boldsymbol{\alpha}_{t}$ and $\mathbf{q}_{t^{-}}$,
find $\mathbf{q}_{t}$) we use a fixed-point argument as described
by Algorithm \ref{algo:q}. Based on our numerical experiment, a stable
solution can be reached within three iterations for a tolerance set
to $\mathrm{tol}=10^{-4}$. This algorithm ensures that the post-transaction
portfolio holdings, accounting for immediate transaction costs, liquidity
costs and market impacts, match exactly the required portfolio allocation
$\alpha_{t}$. Ignoring this actual rebalancing could result in a
large mismatch between the actual post-transaction allocation and
the initial target allocation. We denote the transaction volume as
a function of $\boldsymbol{\alpha}_{t}$, $\mathbf{S}_{t^{-}}$ and
$W_{t^{-}}$, i.e., $\Delta\mathbf{q}_{t}=\mathcal{Q}(\boldsymbol{\alpha}_{t},\mathbf{S}_{t^{-}},W_{t^{-}})$
where $\mathcal{Q}:\mathbb{R}^{d}\times\mathbb{R}^{d}\times\mathbb{R}\rightarrow\mathbb{R}^{d}$.
\begin{algorithm}[H]
\selectlanguage{french}%
\begin{algorithmic}[1]

\selectlanguage{english}%
\STATE \textbf{Input:} $\mathbf{q}_{t^{-}}$, $\mathbf{S}_{t^{-}}$
, $W_{t^{-}}$ and $\boldsymbol{\alpha}_{t}$ 

\vspace{0.2em}

\STATE \textbf{Result:} $\mathbf{q}_{t}$, $q_{t}^{f}$, $\mathbf{S}_{t}$
and $W_{t}$

\vspace{0.1em}

\STATE Set tol

\vspace{0.1em}

\STATE Initial guess: $\mathbf{q}_{t}=\boldsymbol{\alpha}_{t}\times W_{t^{-}}\div\mathbf{S}_{t^{-}}$

\vspace{0.1em}

\WHILE{$\mathrm{dist}>\text{tol}$}

\vspace{0.1em}

\STATE $\mathbf{S}_{t}=\mathbf{S}_{t^{-}}+\textrm{\textbf{MI}}(\Delta\mathbf{q}_{t})$

\vspace{0.2em}

\STATE $W_{t}=W_{t^{-}}-\mathbf{TC}(\Delta\mathbf{q}_{t})\cdot\vec{\mathbf{1}}_{d}-\mathbf{LC}(\Delta\mathbf{q}_{t})\cdot\vec{\mathbf{1}}_{d}+\textrm{\textbf{MI}}(\Delta\mathbf{q}_{t})\cdot\mathbf{q}_{t}$

\vspace{0.2em}

\STATE $\mathbf{q}_{t}^{\mathrm{aux}}=\boldsymbol{\alpha}_{t}\times W_{t}\div\mathbf{S}_{t}$

\vspace{0.2em}

\STATE $\mathrm{dist}=\sum|\mathbf{q}_{t}^{\mathrm{aux}}-\mathbf{q}_{t}|/\mathbf{q}_{t}^{\mathrm{aux}}$

\vspace{0.2em}

\STATE $\mathbf{q}_{t}=\mathbf{q}_{t}^{\mathrm{aux}}$ 

\vspace{0.1em}

\ENDWHILE

\vspace{0.1em}

\STATE $q_{t}^{f}=W_{t}-\mathbf{q}_{t}\cdot\mathbf{S}_{t}$ 

\vspace{0.1em}

\selectlanguage{french}%
\end{algorithmic}

\selectlanguage{english}%
\caption{Compute $\mathbf{q}_{t}$ and $q_{t}^{f}$\label{algo:q}}
\end{algorithm}

The dynamic portfolio allocation is chosen to maximize the investor's
expected utility of final wealth $\mbox{\ensuremath{\mathbb{E}}}\left[U(W_{T})\right]$
over all the possible strategies $\left\{ \boldsymbol{\alpha}_{t}\in\mathcal{A}\right\} _{0\leq t\leq T}$.
Let $\mathcal{F}=\left\{ \mathcal{F}_{t}\right\} _{0\leq t\leq T}$
be the filtration generated by all the state variables. At any time
$t\in[0,T]$, the objective function reads 
\begin{equation}
v_{t}(z,s,w)=\sup_{\left\{ \boldsymbol{\alpha}_{\tau}\in\mathcal{A}\right\} _{t\leq\tau\leq T}}\mbox{\ensuremath{\mathbb{E}}}\left[U(W_{T})\left|\mathbf{Z}_{t}=z,\mathbf{S}_{t^{-}}=s,W_{t^{-}}=w\right.\right],\label{eq:objective-function}
\end{equation}
where $V_{t}=v_{t}(\mathbf{Z}_{t},\mathbf{S}_{t^{-}},W_{t^{-}})$
and $\boldsymbol{\alpha}_{t}$ are $\mathcal{F}_{t}$-adapted. The
state variables of the problem are: 

\begin{enumerate}[leftmargin=*, labelsep=3mm, topsep=0.5mm, itemsep=0mm] 

\item Exogenous state variables: the return predictors $\mathbf{Z}_{t}$

\item Endogenous state variables: the relative portfolio weights
$\boldsymbol{\alpha}_{t}$, the absolute portfolio holdings $\mathbf{q}_{t}$,
the asset prices $\mathbf{S}_{t}$ and the portfolio value $W_{t}$ 

\end{enumerate}

Henceforth, we restrict the rebalancing times to an equally-spaced
discrete grid $0=t_{0}<\cdots<t_{N}=T$. The asset price processes
evolve as
\begin{eqnarray}
\mathbf{S}_{t_{n+1}}\ \  & = & \mathbf{S}_{t_{n}}\times\exp\left(\mathbf{r}_{t_{n+1}}\right)+\mathbf{MI}\left(\Delta\mathbf{q}_{t_{n}}\right),\label{eq:price-evolution}
\end{eqnarray}
and the wealth process evolves as
\begin{equation}
W_{t_{n+1}}=W_{t_{n}}+r^{f}q_{t_{n}}^{f}+\mathbf{q}_{t_{n}}\cdot\left(\mathbf{S}_{t_{n}}\times\mathbf{r}_{t_{n+1}}\right)-\mathbf{TC}\left(\Delta\mathbf{q}_{t_{n+1}}\right)\cdot\vec{\mathbf{1}}_{d}-\mathbf{LC}\left(\Delta\mathbf{q}_{t_{n+1}}\right)\cdot\vec{\mathbf{1}}_{d}+\mathbf{MI}\left(\Delta\mathbf{q}_{t_{n+1}}\right)\cdot\mathbf{q}_{t_{n+1}},\label{eq:wealth-evolution}
\end{equation}
where $\mathbf{q}$ in \eqref{eq:price-evolution}-\eqref{eq:wealth-evolution}
satisfy the relation \eqref{eq:solve-q-alpha}. The value function
satisfies the following discrete dynamic programming principle
\begin{eqnarray}
v_{t_{n}}\left(z,s,w\right) & = & \sup_{\boldsymbol{\alpha}_{t_{n}}\in\mathcal{A}}\mbox{\ensuremath{\mathbb{E}}}\left[v_{t_{n+1}}\left(\mathbf{Z}_{t_{n+1}},\mathbf{S}_{t_{n+1}},W_{t_{n+1}}\right)\left|\mathbf{Z}_{t_{n}}=z,\mathbf{S}_{t_{n}}=s,W_{t_{n}}=w\right.\right]\nonumber \\
v_{t_{N}}\left(z,s,w\right) & = & U(w)\label{eq:dp}
\end{eqnarray}
and we assume that the investor begins with 100\% holding in the cash
account and liquidate all the risky assets at the terminal time, i.e.,
$\boldsymbol{\alpha}_{\left(t_{0}\right)^{-}}=\boldsymbol{\alpha}_{t_{N}}=0$. 

\section{Solution\label{sec:LSMC}}

In this section, we describe our method for solving the recursive
dynamic programming problem \eqref{eq:dp}. Our algorithm can be decomposed
into three main parts: 

\begin{enumerate}[leftmargin=*, labelsep=3mm, topsep=0.5mm, itemsep=0mm] 

\item First, a forward simulation of all the state variables of the
problem, including the endogenous state variables, following the control
randomization method of \citet{Kharroubi2014}, described in Section
\ref{subsec:Simulation};

\item Then, a backward recursive dynamic programming where the conditional
expectations are approximated by least squares regressions, and the
optimal allocation obtained by exhaustive search, described in Section
\ref{subsec:regression-maximization};

\item Finally, an iteration procedure for updating the simulated
control variables of the first step by the estimates generated from
the second step, described in Section \ref{subsec:interpolation}. 

\end{enumerate}

\subsection{Step 1: Monte Carlo simulations\label{subsec:Simulation}}

The first main part consists in simulating a large sample of all the
stochastic state variables. The return predictors $\mathbf{Z}_{t}$
and the asset excess returns $\mathbf{r}_{t}$ are exogenous risk
factors, and therefore easy to simulate. By contrast, the asset prices
$\mathbf{S}_{t}$ and the portfolio value $W_{t}$ are endogenous
risk factors, i.e. their dynamics depend on the control $\boldsymbol{\alpha}_{t}$.
In order to simulate $\mathbf{S}_{t}$, $W_{t}$, which is necessary
to initiate the algorithm, we rely on the control randomization technique
of \citet{Kharroubi2014}. In summary, we first simulate a sample
of return predictors $\left\{ \mathbf{Z}_{t_{n}}^{m}\right\} _{0\leq n\leq N}^{1\leq m\leq M}$,
asset excess returns $\left\{ \mathbf{r}_{t_{n}}^{m}\right\} _{0\leq n\leq N}^{1\leq m\leq M}$
and random portfolio weights $\left\{ \tilde{\boldsymbol{\alpha}}_{t_{n}}^{m}\right\} _{0\leq n\leq N}^{1\leq m\leq M}$,
then compute the corresponding absolute holdings $\left\{ \tilde{\mathbf{q}}_{t_{n}}^{m}\right\} _{0\leq n\leq N}^{1\leq m\leq M}$,
asset prices $\left\{ \tilde{\mathbf{S}}_{\left(t_{n}\right)^{-}}^{m}\right\} _{0\leq n\leq N}^{1\leq m\leq M}$
and portfolio values $\left\{ \tilde{W}_{\left(t_{n}\right)^{-}}^{m}\right\} _{0\leq n\leq N}^{1\leq m\leq M}$
according to Algorithm \ref{algo:q}. The next subsection explains
how these initial random weights will be turned into estimates of
the optimal allocation.

\subsection{Step 2: discretization, regression and maximization\label{subsec:regression-maximization}}

The second part of our LSMC algorithm is the regression and maximization
by exhaustive search. We discretize the control space as $\mathcal{A}\approx\mathcal{A}^{\text{d}}=\{\mathbf{a}_{1},...,\mathbf{a}_{J}\}$.
According to the dynamic programming principle \eqref{eq:dp}, at
time $t_{N}$, the objective function \eqref{eq:objective-function}
is equal to $\hat{v}_{t_{N}}\left(z,s,w\right)=U(w)$. At time $t_{n}$,
assume that the mapping $\hat{v}_{t_{n+1}}:\left(z,s,w\right)\mapsto\hat{v}_{t_{n+1}}\left(z,s,w\right)$
has been estimated, one obtains 
\begin{align*}
 & v_{t_{n}}(z,s,w)\\
= & \sup_{\boldsymbol{\alpha}_{t_{n}}\in\mathcal{A}}\mathbb{E}\left[\left.\hat{v}_{t_{n+1}}\left(\mathbf{Z}_{t_{n+1}},\mathbf{S}_{\left(t_{n+1}\right)^{-}},W_{\left(t_{n+1}\right)^{-}}\right)\right|\mathbf{Z}_{t_{n}}=z,\mathbf{S}_{\left(t_{n}\right)^{-}}=s,W_{\left(t_{n}\right)^{-}}=w\right]\\
\approx & \max_{\mathbf{a}_{j}\in\mathcal{A}^{\text{d}}}\mathbb{E}\left[\left.\hat{v}_{t_{n+1}}\left(\mathbf{Z}_{t_{n+1}},\mathbf{S}_{\left(t_{n+1}\right)^{-}},W_{\left(t_{n+1}\right)^{-}}\right)\right|\mathbf{Z}_{t_{n}}=z,\boldsymbol{\alpha}_{t_{n}}=\mathbf{a}_{j},\mathbf{S}_{\left(t_{n}\right)^{-}}=s,W_{\left(t_{n}\right)^{-}}=w\right].
\end{align*}

By taking the decision $\boldsymbol{\alpha}_{t_{n}}=\mathbf{a}_{j}$,
the endogenous state variables at time $\left(t_{n}\right)^{-}$ can
be updated to their post-transaction values at time $t_{n}$:
\begin{align}
 & v_{t_{n}}(z,s,w)\nonumber \\
= & \max_{\mathbf{a}_{j}\in\mathcal{A}^{\text{d}}}\mathbb{E}\left[\hat{v}_{t_{n+1}}\left(\mathbf{Z}_{t_{n+1}},\mathbf{S}_{\left(t_{n+1}\right)^{-}},W_{\left(t_{n+1}\right)^{-}}\right)\left|\begin{array}{c}
\mathbf{Z}_{t_{n}}=\mathbf{Z}_{t_{n}}^{'},\boldsymbol{\alpha}_{t_{n}}=\boldsymbol{\alpha}_{t_{n}}^{'},\mathbf{q}_{t_{n}}=\mathbf{q}_{t_{n}}^{'},\\
\mathbf{S}_{t_{n}}=\mathbf{S}_{t_{n}}^{'},W_{t_{n}}=W_{t_{n}}^{'}
\end{array}\right.\right]\label{eq:cond_exp}
\end{align}
where
\begin{eqnarray*}
\mathbf{Z}_{t_{n}}^{'} & = & z\\
\boldsymbol{\alpha}_{t_{n}}^{'} & = & \mathbf{a}_{j}\\
\mathbf{q}_{t_{n}}^{'} & = & \mathbf{q}_{t_{n-1}}+\mathcal{Q}\left(\mathbf{a}_{j},s,w\right)\\
\mathbf{S}_{t_{n}}^{'} & = & s+\mathbf{MI}\left(\mathcal{Q}\left(\mathbf{a}_{j},s,w\right)\right)\\
W_{t_{n}}^{'} & = & w-\mathbf{TC}\left(\mathcal{Q}\left(\mathbf{a}_{j},s,w\right)\right)\cdot\vec{\mathbf{1}}_{d}-\mathbf{LC}\left(\mathcal{Q}\left(\mathbf{a}_{j},s,w\right)\right)\cdot\vec{\mathbf{1}}_{d}+\mathbf{MI}\left(\mathcal{Q}\left(\mathbf{a}_{j},s,w\right)\right)\cdot\mathbf{q}_{t_{n}}^{'}
\end{eqnarray*}
Therefore, for each Monte Carlo path $m=1,...,M$, we update the decisions
$\boldsymbol{\alpha}_{t_{n}}^{m}$ to $\mathbf{a}_{j}$ and recompute
the corresponding endogenous variables at time $t_{n}$
\begin{eqnarray*}
\Delta\hat{\mathbf{q}}_{t_{n}}^{m} & = & \mathcal{Q}\left(\mathbf{a}_{j},\tilde{\mathbf{S}}_{\left(t_{n}\right)^{-}}^{m},\tilde{W}_{\left(t_{n}\right)^{-}}^{m}\right)\\
\hat{\mathbf{q}}_{t_{n}}^{m} & = & \mathbf{\tilde{q}}_{t_{n-1}}^{m}+\Delta\hat{\mathbf{q}}_{t_{n}}^{m}\\
\hat{\mathbf{S}}_{t_{n}}^{m} & = & \tilde{\mathbf{S}}_{\left(t_{n}\right)^{-}}^{m}+\mathbf{MI}\left(\Delta\hat{\mathbf{q}}_{t_{n}}^{m}\right)\\
\hat{W}_{t_{n}}^{m} & = & \tilde{W}_{\left(t_{n}\right)^{-}}^{m}-\mathbf{TC}\left(\Delta\hat{\mathbf{q}}_{t_{n}}^{m}\right)\cdot\vec{\mathbf{1}}_{d}-\mathbf{LC}\left(\Delta\hat{\mathbf{q}}_{t_{n}}^{m}\right)\cdot\vec{\mathbf{1}}_{d}+\mathbf{MI}\left(\Delta\hat{\mathbf{q}}_{t_{n}}^{m}\right)\cdot\hat{\mathbf{q}}_{t_{n}}^{m},
\end{eqnarray*}
then recompute the endogenous state variables one time-step forward
at time $t_{\left(n+1\right)^{-}}$, i.e.,
\begin{eqnarray*}
\hat{\mathbf{S}}_{\left(t_{n+1}\right)^{-}}^{m} & = & \hat{\mathbf{S}}_{t_{n}}^{m}\times\exp\left(\mathbf{r}_{t_{n+1}}^{m}\right)\\
\hat{W}_{\left(t_{n+1}\right)^{-}}^{m} & = & \hat{W}_{t_{n}}^{m}+r^{f}q_{t_{n}}^{f,m}+\hat{\mathbf{q}}_{t_{n}}^{m}\cdot\left(\hat{\mathbf{S}}_{t_{n}}^{m}\times\mathbf{r}_{t_{n+1}}^{m}\right).
\end{eqnarray*}
Finally, set $\left\{ L_{k}(z,s,w)\right\} _{1\leq k\leq K}$ to be
a vector of basis functions of state variables. We estimate the ``continuation
values'' (the conditional expectations in equation \eqref{eq:cond_exp})
by least squares minimization, i.e.,
\begin{eqnarray}
\left\{ \hat{\beta}_{k,t_{n}}^{j}\right\} _{1\leq k\leq K} & = & {\displaystyle \arg\min_{\beta\in\mathbb{R}^{K}}}\sum_{m=1}^{M}\left(\begin{array}{c}
\sum_{k=1}^{K}\beta_{k}L_{k}\left(\left.\mathbf{Z}_{t_{n}}^{m},\hat{\mathbf{S}}_{t_{n}}^{m},\hat{W}_{t_{n}}^{m}\right|\boldsymbol{\alpha}_{t_{n}}^{m}=\mathbf{a}_{j}\right)\\
-\hat{v}_{t_{n+1}}\left(\left.\mathbf{Z}_{t_{n+1}}^{m},\hat{\mathbf{S}}_{\left(t_{n+1}\right)^{-}}^{m},\hat{W}_{\left(t_{n+1}\right)^{-}}^{m}\right|\boldsymbol{\alpha}_{t_{n}}^{m}=\mathbf{a}_{j}\right)
\end{array}\right)^{2}.\label{eq:ls}
\end{eqnarray}
Therefore the ``continuation value'' at time $t_{n}$ for $\mathbf{a}_{j}\in\mathcal{A}^{\text{d}}$
is formulated as
\begin{eqnarray*}
\hat{\text{CV}}_{t_{n}}^{j}\left(z,s,w\right) & = & \sum_{k=1}^{K}\hat{\beta}_{k,t_{n}}^{j}L_{k}\left(z,s,w\right),
\end{eqnarray*}
and the mappings $\hat{\boldsymbol{\alpha}}_{t_{n}}:\left(z,s,w\right)\mapsto\hat{\boldsymbol{\alpha}}_{t_{n}}\left(z,s,w\right)$
and $\hat{v}_{t_{n}}:\left(z,s,w\right)\mapsto\hat{v}_{t_{n}}\left(z,s,w\right)$
are estimated by
\begin{eqnarray*}
\hat{\boldsymbol{\alpha}}_{t_{n}}\left(z,s,w\right)=\arg\max_{\mathbf{a}_{j}\in\mathcal{A}^{\text{d}}}\hat{\text{CV}}{}_{t_{n}}^{j}\left(z,s,w\right) & \text{ or } & \hat{v}_{t_{n}}(z,s,w)=\max_{\mathbf{a}_{j}\in\mathcal{A}^{\text{d}}}\hat{\text{CV}}{}_{t_{n}}^{j}\left(z,s,w\right).
\end{eqnarray*}
It is important to remark that the discretization of the control allowed
us to substitute the extended control regression of \citet{Kharroubi2014}
by one regression \eqref{eq:ls} for each control level. 

\subsection{Step 3: control iteration\label{subsec:interpolation}}

In the forward simulation, the endogenous state variables are generated
using the randomized controls $\left\{ \tilde{\boldsymbol{\alpha}}_{t_{n}}^{m}\right\} _{0\leq n\leq N}^{1\leq m\leq M}$.
Although the endogenous state variables will be updated and corrected
backwards during Step 2, the evaluation of
\[
v_{t_{n}}(z,s,w)=\sup_{\boldsymbol{\alpha}_{t_{n}}\in\mathcal{A}}\mathbb{E}\left[\left.\hat{v}_{t_{n+1}}\left(\mathbf{Z}_{t_{n+1}},\mathbf{S}_{\left(t_{n+1}\right)^{-}},W_{\left(t_{n+1}\right)^{-}}\right)\right|\mathbf{Z}_{t_{n}}=z,\mathbf{S}_{\left(t_{n}\right)^{-}}=s,W_{\left(t_{n}\right)^{-}}=w\right]
\]
is made on the sample of path-dependent variables $\left\{ \mathbf{S}_{\left(t_{n}\right)^{-}}^{m},W_{\left(t_{n}\right)^{-}}^{m}\right\} _{1\leq m\leq M}$
which still depend on the historical randomized controls $\left\{ \tilde{\boldsymbol{\alpha}}_{t_{n}}^{m}\right\} _{0\leq n'\leq n-1}^{1\leq m\leq M}$.
In theory, this fact does not affect the optimality of the allocation
estimates, as the regression provides an estimate of $v_{t_{n}}(z,s,w)$
everywhere, including the region where the optimally controlled endogenous
variables $\mathbf{S}_{\left(t_{n}\right)^{-}}$ and $W_{\left(t_{n}\right)^{-}}$
will eventually lie. In practice, it may lead to possibly large numerical
errors if the regression is numerically inaccurate in the optimal
region, due to an insufficiently large sample size or inadequate regression
basis for example. To mitigate this possibility, we propose to iterate
the whole algorithm, with the initial randomized controls replaced
by the estimated optimal controls produced by the previous run. This
iteration procedure will bring the whole sample $\left\{ \mathbf{S}_{\left(t_{n}\right)^{-}}^{m},W_{\left(t_{n}\right)^{-}}^{m}\right\} _{1\leq m\leq M}$
closer to the optimal region, and thus improve the overall portfolio
allocation estimates. Our numerical experiments in Section \ref{sec:Application}
show that this iteration procedure does improve accuracy, especially
for small sample sizes and highly nonlinear utility functions, and
that most of the improvements occur after one single additional iteration. 

\subsection{Summary and remarks}

Finally, this subsection provides a detailed description of the backward
iterations, followed by a few additional implementation details.

\paragraph{Summary of algorithm}

Being given the ``continuation values'' at time $t_{N-1}$, the
detailed implementation of one backward iteration (cf. Section \ref{subsec:regression-maximization})
is summarized in Algorithm \ref{algo:lsmc}, where we set $\left\{ \tilde{\boldsymbol{\alpha}}_{\left(t_{0}\right)^{-}}^{m}\right\} _{1\leq m\leq M}=0$,
$\left\{ \tilde{\mathbf{q}}_{\left(t_{0}\right)^{-}}^{m}\right\} _{1\leq m\leq M}=0$
and $\tilde{q}_{\left(t_{0}\right)^{-}}^{f,m}=W_{\left(t_{0}\right)^{-}}=\text{initial investment amount}$.
Additional implementation details are discussed below.

\begin{algorithm}[H]
\begin{algorithmic}[1]

\vspace{0.1em}

\STATE \textbf{Input:} $\left(\mathbf{Z}_{t_{n}}^{m},\mathbf{r}_{t_{n}}^{m},\tilde{\boldsymbol{\alpha}}_{t_{n}}^{m},\tilde{\mathbf{q}}_{t_{n}}^{m},\tilde{\mathbf{S}}_{\left(t_{n}\right)^{-}}^{m},\tilde{W}_{\left(t_{n}\right)^{-}}^{m}\right){}_{0\leq n\leq N}^{1\leq m\leq M}$,
$\left(\hat{\text{CV}}_{t_{N-1}}^{j},\hat{\beta}_{t_{N-1}}^{j}\right)_{1\leq j\leq J}$ 

\vspace{0.1em}

\STATE \textbf{Result:} $\hat{\boldsymbol{\alpha}}_{t_{0}}$ 

\vspace{0.1em}

\FORALL{rebalancing time $t_{n}=t_{N-1},\ldots,t_{0}$}

\vspace{0.1em}

\FORALL{decision $\mathbf{a}_{j}\in\mathcal{A}^{\text{d}}$}

\vspace{0.1em}

\FORALL{Monte Carlo path $m=1,...,M$}

\vspace{0.1em}

\STATE Compute $\left(\hat{\mathbf{q}}_{t_{n}}^{m},\hat{\mathbf{S}}_{t_{n}}^{m},\hat{W}_{t_{n}}^{m}\right)$
from $\left(\tilde{\mathbf{q}}_{t_{n-1}}^{m},\tilde{\mathbf{S}}_{\left(t_{n}\right)^{-}}^{m},\tilde{W}_{\left(t_{n}\right)^{-}}^{m},\boldsymbol{\alpha}_{t_{n}}^{m}=\mathbf{a}_{j}\right)$
using Algorithm \ref{algo:q}

\vspace{0.2em}

\STATE Compute $\hat{\mathbf{S}}_{\left(t_{n+1}\right)^{-}}^{m}=\hat{\mathbf{S}}_{t_{n}}^{m}\times\exp\left(\mathbf{r}_{t_{n+1}}^{m}\right)$
and $\hat{W}_{\left(t_{n+1}\right)^{-}}^{m}=\hat{W}_{t_{n}}^{m}+r^{f}q_{t_{n}}^{f,m}+\hat{\mathbf{q}}_{t_{n}}^{m}\cdot\left(\hat{\mathbf{S}}_{t_{n}}^{m}\times\mathbf{r}_{t_{n+1}}^{m}\right)$

\vspace{0.1em}

\FORALL{rebalancing time $t_{n'}=t_{n+1},\ldots,t_{N-1}$}

\vspace{0.1em}

\FORALL{decision $\mathbf{a}_{l}\in\mathcal{A}^{\text{d}}$}

\vspace{0.1em}

\STATE Compute $\left(\hat{\mathbf{q}}_{t_{n'}}^{m},\hat{\mathbf{S}}_{t_{n'}}^{m},\hat{W}_{t_{n'}}^{m}\right)$
from $\left(\hat{\mathbf{q}}_{t_{n'-1}}^{m},\hat{\mathbf{S}}_{\left(t_{n'}\right)^{-}}^{m},\hat{W}_{\left(t_{n'}\right)^{-}}^{m},\boldsymbol{\alpha}_{t_{n'}}=\mathbf{a}_{l}\right)$
using Algorithm \ref{algo:q}

\vspace{0.2em}

\STATE Compute $\hat{\text{CV}}_{t_{n'}}^{l}\left(\mathbf{Z}_{t_{n'}}^{m},\hat{\mathbf{S}}_{t_{n'}}^{m},\hat{W}_{t_{n'}}^{m}\right)=\sum_{k=1}^{K}\hat{\beta}_{k,t_{n'}}^{l}L_{k}\left(\left.\mathbf{Z}_{t_{n'}}^{m},\hat{\mathbf{S}}_{t_{n'}}^{m},\hat{W}_{t_{n'}}^{m}\right|\boldsymbol{\alpha}_{t_{n'}}^{m}=\mathbf{a}_{l}\right)$

\vspace{0.1em}

\ENDFOR

\vspace{0.1em}

\STATE Update $\left(\hat{\mathbf{q}}_{t_{n'}}^{m},\hat{\mathbf{S}}_{t_{n'}}^{m},\hat{W}_{t_{n'}}^{m}\right)$
with $\boldsymbol{\alpha}_{t_{n'}}={\displaystyle \arg\max_{\mathbf{a}_{l}\in\mathcal{A}^{\text{d}}}}\hat{\text{CV}}_{t_{n'}}^{l}\left(\mathbf{Z}_{t_{n'}}^{m},\hat{\mathbf{S}}_{t_{n'}}^{m},\hat{W}_{t_{n'}}^{m}\right)$ 

\vspace{0.2em}

\STATE Compute $\hat{\mathbf{S}}_{\left(t_{n'+1}\right)^{-}}^{m}$=$\hat{\mathbf{S}}_{t_{n'}}^{m}\times\exp\left(\mathbf{r}_{t_{n'+1}}^{m}\right)$
and $\hat{W}_{\left(t_{n'+1}\right)^{-}}^{m}$=$\hat{W}_{t_{n'}}^{m}+r^{f}q_{t_{n'}}^{f,m}+\hat{\mathbf{q}}_{t_{n'}}^{m}\cdot\left(\hat{\mathbf{S}}_{t_{n'}}^{m}\times\mathbf{r}_{t_{n'+1}}^{m}\right)$

\vspace{0.1em}

\ENDFOR

\vspace{0.1em}

\STATE Compute $\hat{W}_{t_{N}}^{m}$ from $\left(\hat{\mathbf{q}}_{t_{N-1}}^{m},\hat{\mathbf{S}}_{\left(t_{N}\right)^{-}}^{m},\hat{W}_{\left(t_{N}\right)^{-}}^{m},\boldsymbol{\alpha}_{t_{N}}^{m}=0\right)$
using Algorithm \ref{algo:q}

\vspace{0.1em}

\ENDFOR

\vspace{0.1em}

\IF{$t_{n}>t_{0}$}

\vspace{0.1em}

\STATE Least-squares approximation with basis functions of state
variables, $\left\{ L_{k}\left(z,s,w\right)\right\} _{1\leq k\leq K}$:\vspace{-1em}
\[
\left\{ \hat{\beta}_{k,t_{n}}^{j}\right\} _{1\leq k\leq K}={\displaystyle \arg\min_{\beta\in\mathbb{R}^{K}}}\sum_{m=1}^{M}\left(\sum_{k=1}^{K}\beta_{k}L_{k}\left(\left.\mathbf{Z}_{t_{n}}^{m},\hat{\mathbf{S}}_{t_{n}}^{m},\hat{W}_{t_{n}}^{m}\right|\boldsymbol{\alpha}_{t_{n}}^{m}=\mathbf{a}_{j}\right)-U\left(\hat{W}_{t_{N}}^{m}\right)\right)^{2}
\]
\vspace{-1em}

\STATE Formulate: $\hat{\text{CV}}{}_{t_{n}}^{j}\left(z,s,w\right)=\sum_{k=1}^{K}\hat{\beta}_{k,t_{n}}^{j}\cdot L_{k}\left(z,s,w\right)$

\vspace{0.1em}

\ELSE

\vspace{0.1em}

\STATE Compute: $\hat{\text{CV}}{}_{t_{0}}^{j}=\frac{1}{M}\sum_{m=1}^{M}U\left(\hat{W}_{t_{N}}^{m}\right)$

\vspace{0.1em}

\ENDIF

\vspace{0.1em}

\ENDFOR

\vspace{0.1em}

\ENDFOR

\vspace{0.1em}

\STATE Initial optimal control: $\hat{\boldsymbol{\alpha}}_{t_{0}}=\arg{\displaystyle \max_{\mathbf{a}_{j}\in\mathcal{A}^{\text{d}}}}\hat{\text{CV}}_{t_{0}}^{j}$ 

\end{algorithmic}

\caption{Backward Dynamic Programming\label{algo:lsmc}}
\end{algorithm}

\paragraph{VFI versus PFI}

Two alternative implementations of the LSMC algorithm can be used:
value function iteration (VFI, \citet{Carriere1996}, \citet{Tsitsiklis01},
a.k.a. regression surface value iteration), and performance function
iteration (PFI, \citet{Longstaff2001}, a.k.a. realized value iteration,
or portfolio weight iteration). The difference lies in the $t_{n+1}-$response
in the least squares regressions \eqref{eq:ls}: the VFI scheme regresses
the estimated continuation value function from the previous regression,
while the PFI scheme regresses the realized paths under the estimated
optimal policy. The PFI scheme produces more accurate results, as
it avoids the compounding of regression errors of the VFI scheme.
However, when some state variables are endogenous, the PFI scheme
requires to recompute all the endogenous state variables until the
end of the horizon, which increases the computational complexity from
linear to quadratic in time. By contrast, the computational complexity
of the VFI is linear in the number of time steps. More discussions
on VFI versus PFI are available in \citet{vanBinsbergen2007}, \citet{Garlappi2009}
and \citet{Denault2017}. In this paper, we choose to implement the
PFI scheme for its greater accuracy and stability.

\paragraph*{Dimension reduction of state vector}

Although in theory all the risk factors need to be included in the
regression so as to take all the available information into account
when making decisions, in practice the bias-variance tradeoff suggests
to omit the variables that bring little additional information. In
portfolio allocation problems, the portfolio wealth is a linear combination
of the asset prices, determined by $W_{t_{n}}=\mathbf{S}_{t_{n}}\cdot\mathbf{q}_{t_{n}}$,
such that most of the relevant price changes can be reflected in a
single wealth variable. Moreover, our objective is to maximize the
expected utility of final wealth, thus the wealth variable plays a
much more crucial role when approximating such objective function
than the price variables. After testing and comparing different subsets
of regression inputs, we decided to remove the endogenous price variables
in the regressions and only regress on $\left(\mathbf{Z},W\right)$.
Doing so improves the out-of-sample quality of regression estimates,
and has the advantage that the number of assets does not increase
the numerical complexity of each least-squares regression in the LSMC
algorithm.

\paragraph*{Regressing on post- versus pre-transaction variables}

The evolution of the endogenous state variables from time $t_{n^{-}}$
to $t_{N}$ can be decomposed into an immediate deterministic component
depending on the switching costs $\mathbf{TC}\left(\Delta\mathbf{q}_{t_{n}}\right)$,
$\mathbf{LC}\left(\Delta\mathbf{q}_{t_{n}}\right)$ and $\mathbf{MI}\left(\Delta\mathbf{q}_{t_{n}}\right)$,
and a stochastic component depending on the dynamics of the state
variables from time $t_{n}$ to $t_{N}$. For demonstration purposes,
we use in this paragraph the wealth variable $W$ as one single regressor
in the regression, use a simple linear utility function $\mathcal{U}(w)=w$,
denote $\text{SC}\left(\Delta\mathbf{q}_{t_{n}}\right)$ as the corresponding
overall switching cost which is the immediate deterministic component
at time $t_{n}$, and denote the stochastic component evolving from
time $t_{n}$ to $t_{N}$ as $\Delta{}_{t_{n},t_{N}}$. Then the two
alternative regressions are given by 
\begin{eqnarray}
\text{regression on }W_{\left(t_{n}\right)^{-}}: &  & \mathbb{E}\left[W_{\left(t_{n}\right)^{-}}-\text{SC}\left(\Delta\mathbf{q}_{t_{n}}\right)+\Delta{}_{t_{n},t_{N}}\left|W_{\left(t_{n}\right)^{-}}\right.\right]\approx\beta W_{\left(t_{n}\right)^{-}}\label{eq:pre-transaction}\\
\text{regression on }W_{t_{n}}: &  & \mathbb{E}\left[W_{\left(t_{n}\right)^{-}}-\text{SC}\left(\Delta\mathbf{q}_{t_{n}}\right)+\Delta{}_{t_{n},t_{N}}\left|W_{t_{n}}\right.\right]\approx\beta\left(W_{\left(t_{n}\right)^{-}}-\text{SC}\left(\Delta\mathbf{q}_{t_{n}}\right)\right)\label{eq:post-transaction}
\end{eqnarray}
Here, the pre-transaction regression \eqref{eq:pre-transaction} accounts
for both the deterministic and stochastic evolutions, while the post-transaction
regression \eqref{eq:post-transaction} accounts for the stochastic
evolution only. We favor the regression on post-transaction variables
for several reasons. Firstly, the deterministic component $\text{SC}\left(\Delta\mathbf{q}_{t_{n}}\right)$
is $\mathcal{F}_{t_{n}}$-adapted, and thus not necessary for the
regression. Secondly, the switching costs $\text{SC}\left(\Delta\mathbf{q}_{t_{n}}\right)$
are, at this stage of the algorithm, computed from randomized portfolio
positions $\tilde{\mathbf{q}}_{t_{n-1}}$, and thus also randomized.
Consequently, the switching costs $\left\{ \text{SC}\left(\Delta\mathbf{q}_{t_{n}}^{m}\right)\right\} _{1\leq m\leq M}$
are not smooth w.r.t the regressor $\left\{ W_{\left(t_{n}\right)^{-}}^{m}\right\} _{1\leq m\leq M}$,
which may lead to a substantial information loss w.r.t. the switching
cost by anchoring the unsmoothness around $\mathbb{E}\left[\left.\text{SC}\left(\Delta\mathbf{q}_{t_{n}}\right)\right|W_{\left(t_{n}\right)^{-}}\right]$
(overestimation of the conditional expectation \eqref{eq:pre-transaction}
for large $\text{SC}\left(\Delta\mathbf{q}_{t_{n}}\right)$ realizations,
and underestimation of the conditional expectation \eqref{eq:pre-transaction}
for small $\text{SC}\left(\Delta\mathbf{q}_{t_{n}}\right)$ realizations).
Therefore, subtracting the switching costs from the regressor will
avoid this problem by removing this auxiliary randomness from the
regression. 

Finally, from a practical decision point of view, an investor would
consider the known, immediate transaction cost, liquidity cost and
market impact when making a portfolio rebalancing decision.

\section{Power law liquidity function \label{sec:liquidity}}

One key feature of the presented portfolio allocation algorithm is
its flexibility to accommodate general transaction cost, liquidity
cost and market impact. This is the reason why the presented algorithm
has so far involved general costs $\mathbf{TC}$, $\mathbf{LC}$ and
$\mathbf{MI}$. In this section, we now specify a realistic model
for these costs in view of implementation and testing in the next
Section \ref{sec:Application}.

Transaction cost refers to the commission fee charged by the broker,
usually a fixed amount or a fixed proportional rate, and therefore
easy to account for. The focus of the paper will be liquidity cost
and market impact. During a transaction, the following are the key
observables:
\begin{eqnarray*}
S_{t^{-}} & = & \text{market price before the transaction begins}\\
S_{t}\  & = & \text{market price immediately after the transaction is completed}\\
\bar{S}_{t}\  & = & \text{trading volume-weighted average price on the transaction}
\end{eqnarray*}
In our framework, the post-transaction price $S_{t}$ captures the
(permanent) market impact, i.e., $\text{MI}=S_{t}-S_{t^{-}}$ and
the average price captures the (temporary) liquidity cost, i.e., $\text{LC}=\left|\bar{S}_{t}-S_{t^{-}}\right|$.

In reality, the shape of the limit order book differs by the characteristics
of the portfolio assets. A power law of both liquidity cost and market
impact for the U.S. stock markets has been found in \citet{Almgren2005}.
\citet{Obizhaeva2013} assume a linear price shift and uses a negative
exponential function to model the resilience of the limit order book.
\citet*{Tian2013} found a `square-root' relation between the price
and the available market orders and for large or medium-cap equities
and a `square' relation for small-cap equities in the European market.
A different type of `square-root' relation is shown in \citet*{Cont2014}
for the stocks listed on NYSE. 

In this paper, we adopt the calibrated power law functions of \citet{Almgren2005}
to analyze the impact of the market illiquidity on the dynamic portfolio
selection problem. These power law functions are given by
\begin{eqnarray}
\text{MI}(\Delta q) & = & 0.314\cdot\sigma_{\text{day}}\cdot\frac{\Delta q}{\text{Vol}_{\text{day}}}\cdot\left(\frac{\Theta}{\text{Vol}_{\text{day}}}\right)^{1/4}\label{eq:impact-func}\\
\text{LC}(\Delta q) & = & \left|\frac{\text{MI}(\Delta q)}{2}+0.142\cdot\text{sign}(\Delta q)\cdot\sigma_{\text{day}}\cdot\left|\frac{\Delta q}{\delta\cdot\text{Vol}_{\text{day}}}\right|^{3/5}\right|\label{eq:liquidity-func}
\end{eqnarray}
where $\text{Vol}^{\text{day}}$ is the daily trading volume of the
stock, $\sigma^{\text{day}}$ is the daily volatility of the stock
price, $\delta$ is the time length of trade execution, $\Theta$
is the number of outstanding shares. In the following numerical section,
we will fix the values of $\delta$ and $\Theta$ and focus on the
impact of $\sigma^{\text{day}}$ and $\text{Vol}^{\text{day}}$ on
the portfolio selection problem. 

\section{Numerical experiments\label{sec:Application}}

In this section, we test our algorithm on a cash and stock portfolio.
The outline of this numerical section is as follows:

\begin{enumerate}[leftmargin=*, labelsep=3mm, topsep=0.5mm, itemsep=0mm] 

\item Subection \ref{subsec:Monte-Carlo-convergence} validates the
Monte Carlo convergence of our method with different risk aversion
levels, investment horizon and liquidity settings.

\item Subection \ref{subsec:evolution} discusses the time evolution
of the distribution of portfolio value and the percentage allocation
under different liquidity settings.

\item Subection \ref{subsec:CER-losses} identifies the certainty
equivalent losses associated with ignoring liquidity effects.

\item Finally, subection \ref{subsec:Sensitivity} provides sensitivity
analyzes of the portfolio performance and allocation with respect
to liquidity settings.

\end{enumerate}

But first, we detail the numerical settings used to perform the numerical
experiments reported in this section.

\paragraph{Data and modeling}

Table \ref{tab:assets} summarizes the financial instruments considered
for return predictors. We calibrate a first order vector autoregressive
model to monthly log-returns (i.e., $\log S_{t}-\log S_{t-1}$) from
October 2007 to January 2016\footnote{These data are obtained from Yahoo Finance.}.
We assume the annual interest rate on the cash account is $1.2\%$
and use SPDR S\&P500 index ETF as the proxy for the stock return.

\paragraph*{Switching costs}

To focus on the liquidity effects, we assume for simplicity no fixed
or proportional transaction cost in the numerical study. Regarding
liquidity cost and market impact modeling, we use the power law functions
\eqref{eq:liquidity-func}, where we assume the number of outstanding
share $\Theta=988m$ and the trading duration $\delta=5\min$. We
will analyze the liquidity effects characterized by different levels
of ($\sigma_{\text{day}},\text{Vol}_{\text{day}}$) and we follow
the usual U.S. equity markets such that $\sigma_{\text{day}}\in[2,13]$
and $\text{Vol}_{\text{day}}\in[10m,120m]$.

\paragraph{Certainty equivalent return}

For all the numerical tests, we report the portfolio performances
in terms of monthly adjusted certainty equivalent returns (CER) calculated
by
\[
\mathrm{CER}=U^{-1}\left(\mathbb{E}\left[U(W_{T})\right]\right)^{\frac{1}{T}}-1\approx U^{-1}\left(\frac{1}{M}\sum_{m=1}^{M}U(W_{T}^{m})\right)^{\frac{1}{T}}-1.
\]

The magnitude of monthly returns is usually less than one percent,
thus we display the certainty equivalent returns in basis points ($0.01\%$)
to make comparisons easier. 

\paragraph*{LSMC settings}

We use $M=10^{5}$ Monte Carlo simulations and $N=12$ monthly time
steps (one year horizon and $12$ rebalancing periods), except when
we test the numerical sensitivity to these two parameters (subsection
\ref{subsec:Monte-Carlo-convergence}). After the LSMC algorithm is
completed, we generate another sample of $M=10^{5}$ to calculate
the CER. We denote $I$ as the number of additional control iterations
of the whole LSMC algorithm (subsection \ref{subsec:interpolation}),
$I=0$ meaning only one LSMC run and no additional iterations.

\paragraph{Portfolio weight}

We denote $\alpha$ as the percentage allocation to the stock component,
and $1-\alpha$ as the allocation to the cash component. We assume
a discrete set of admissible controls with step size $0.01$, i.e.,
$\alpha\in\{0.01,0.02,...0.99,1.00\}=\mathcal{A}^{\text{d}}$.

\paragraph{Basis function and regression }

We first scale all the exogenous risk factors (in our case the log-returns)
by dividing by their unconditional mean. For the endogenous risk factor
(the portfolio wealth $W$), we transform it as $U(W/W_{0})$, where
$W_{0}$ is the initial portfolio wealth and $U(\cdot)$ is the CRRA
utility function. These transformed quantities form the inputs of
our regression basis. For the regression basis, we use a simple second
order multivariate polynomial basis. We chose this basis and its order
by observing the plots of the objective function w.r.t. the regression
bases at various intermediate times. The surface shape was found to
be close to linear but slightly curved, suggesting that polynomials
of order two could be sufficient. 

\subsection{Monte Carlo convergence \label{subsec:Monte-Carlo-convergence}}

Table \ref{tab:mc-benchmark} reports the Monte Carlo convergence
of the portfolio allocation algorithm \ref{algo:lsmc} described in
Section \ref{sec:LSMC} on a simple cash and stock allocation problem
with CARA utility $U(w)=-\exp(-\gamma w)$, risk-free rate $0.012$,
and a stock annual return with mean $0.03$ and volatility $0.15$.
These convergence results are compared to the original control randomization
algorithm of \citet{Kharroubi2014} (KLP) for which we include the
portfolio allocation into the same second-order global polynomial
basis. The main observation is that Algorithm \ref{algo:lsmc} uniformly
improves the accuracy of the KLP algorithm with second-order basis.
The improvement is more substantial with long maturities ($N=15$),
large risk-aversion ($\gamma=15$) or small sample size ($M=10^{3}$).
In these three cases, the benefit of using control iteration (subsection
\ref{subsec:interpolation}) is noticeable, and most of the improvement
is achieved after one single additional control iteration ($I=1$).

A similar result can be observed in Table \ref{tab:mc-liquidity}
where the Monte Carlo convergence of CER is reported for different
liquidity settings characterized by daily volatility $\sigma_{\text{day}}$
and daily trading volume $\text{Vol}_{\text{day}}$. Once again, Algorithm
\ref{algo:lsmc} with one additional control iteration ($I=1$) is
superior to both KLP and Algorithm \ref{algo:lsmc} with no additional
iteration ($I=0$). Adding further control iterations ($I=2$ and
more) does not bring significant improvement over $I=1$. When the
market liquidity effects are small, e.g., small $\sigma_{\text{day}}$
or large $\text{Vol}_{\text{day}}$, a small Monte Carlo sample size
is enough for convergence, while a large sample size is needed for
large market liquidity effects, e.g., large $\sigma_{\text{day}}$
or small $\text{Vol}_{\text{day}}$. For the rest of this numerical
section, we will use $M=10^{5}$ with $I=1$ to ensure convergence
and accuracy.

A final remark is that, for large enough sample size ($M\geq10^{4}$),
our LSMC method with $I=0$ greatly outperforms the KLP algorithm
for large risk-aversion levels (Table \ref{tab:mc-benchmark}), but
makes little difference for large switching costs but low risk-aversion
levels (Table \ref{tab:mc-liquidity}), indicating that the nonlinearity
of the final payoff function plays a more crucial role than the size
of switching costs in the accuracy of simulation-and-regression approximating
dynamic programming schemes.

\subsection{Time-evolution of distribution of control and wealth\label{subsec:evolution}}

\begin{figure}[H]
\caption{Time evolution of the distribution of the control and wealth\label{fig:liquidity-control}}

\smallskip
\begin{centering}
\hspace{-3em}%
\begin{minipage}[t]{0.4\columnwidth}%
\includegraphics[scale=0.14]{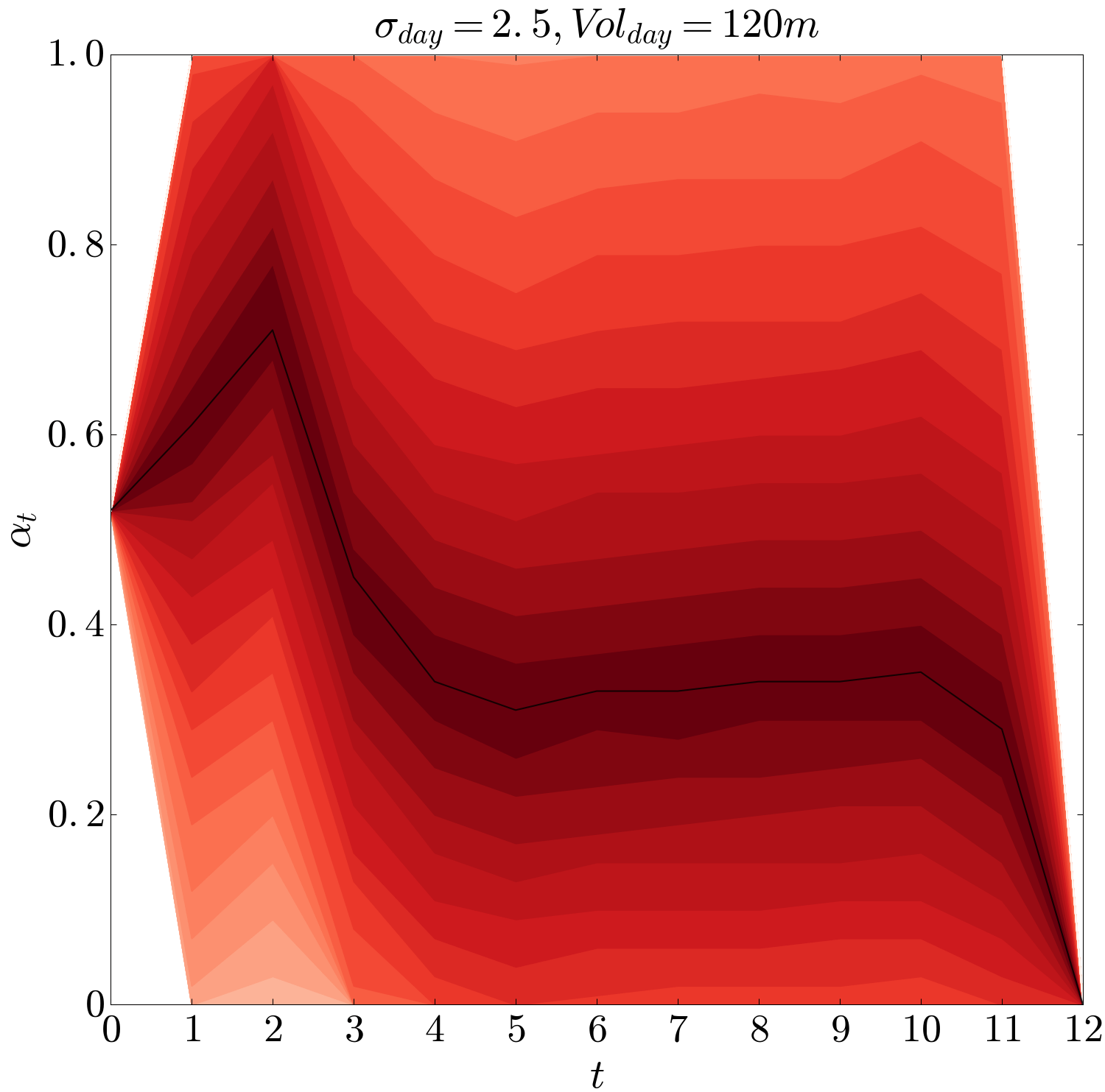}%
\end{minipage}\hspace{3em}%
\begin{minipage}[t]{0.4\columnwidth}%
\includegraphics[scale=0.14]{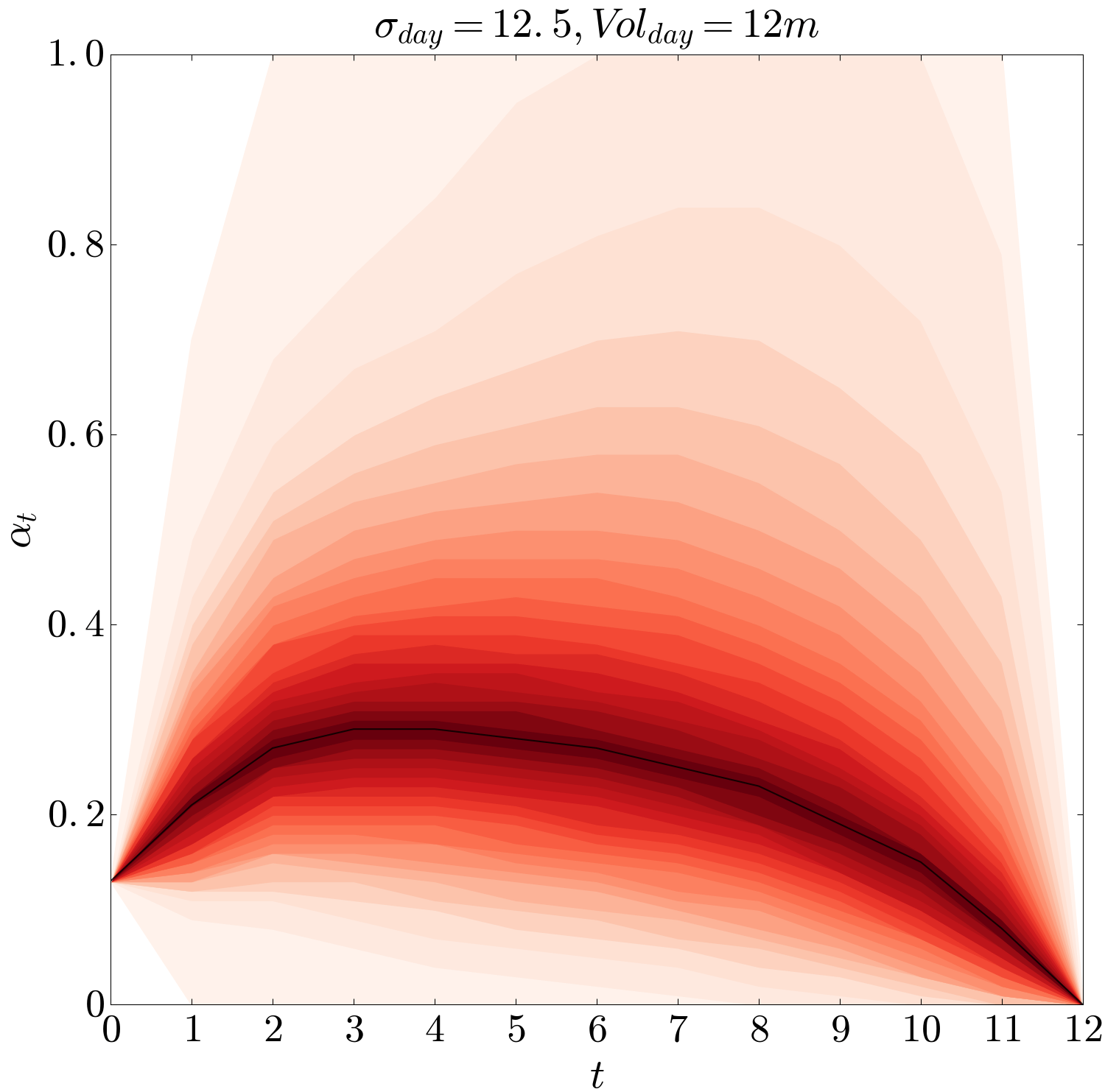}%
\end{minipage}
\par\end{centering}
\begin{centering}
\hspace{-3em}%
\begin{minipage}[t]{0.4\columnwidth}%
\includegraphics[scale=0.14]{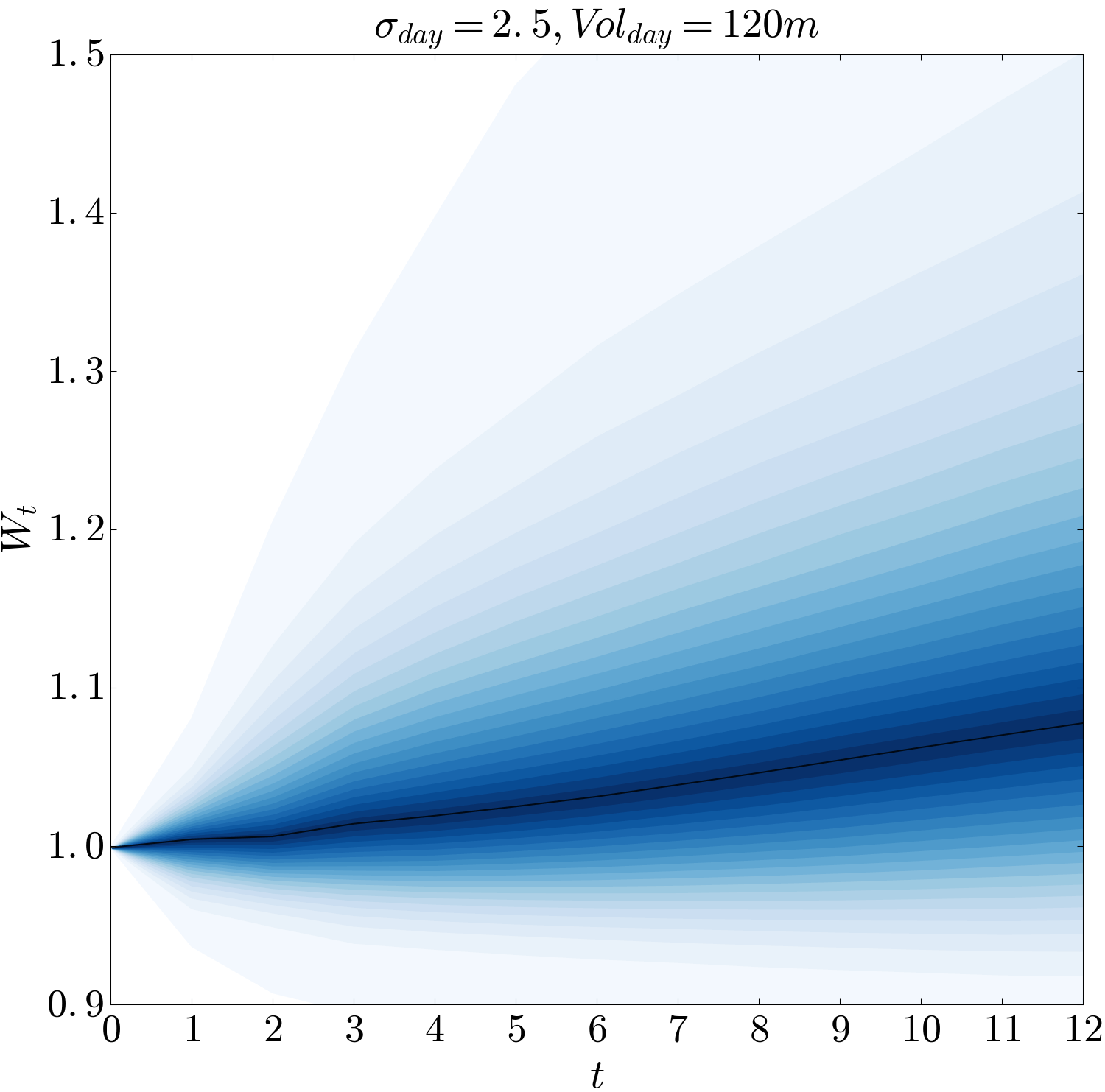}%
\end{minipage}\hspace{3em}%
\begin{minipage}[t]{0.4\columnwidth}%
\includegraphics[scale=0.14]{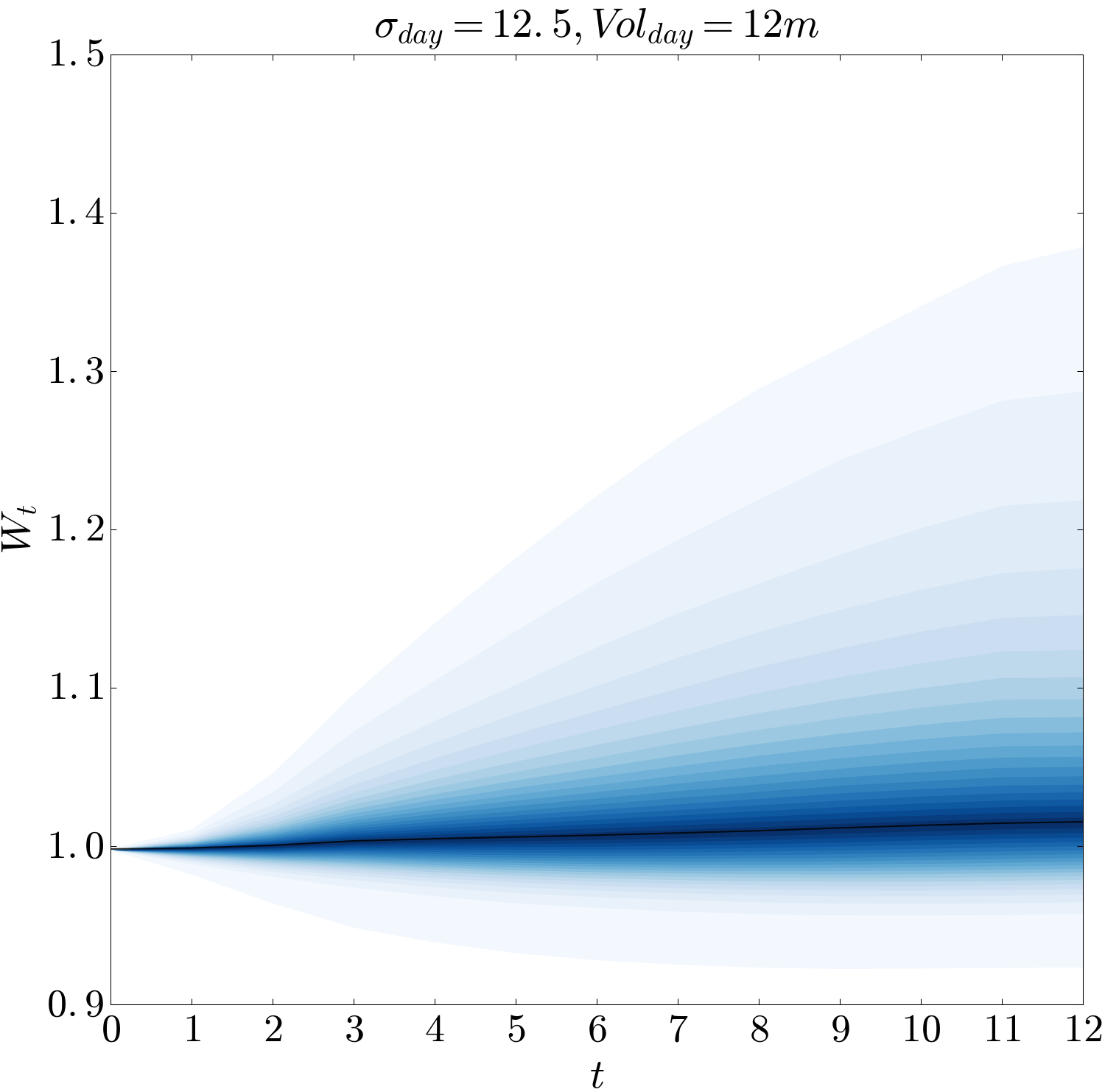}%
\end{minipage}
\par\end{centering}
\smallskip

\begin{singlespace}
{\footnotesize{}This figure shows the time evolution of the distribution
of the }\textbf{\footnotesize{}portfolio allocation (top panel)}{\footnotesize{}
and }\textbf{\footnotesize{}wealth (bottom panel)}{\footnotesize{}
for a CRRA utility with $\gamma=5$ and investment horizon $N=12$
months under different liquidity effects $(\sigma_{\text{day}},\text{Vol}_{\text{day}})=\left(2.5,120m\right),(12.5,12m)$,
using $M=10^{5}$ Monte Carlo simulations with one control iteration
$I=1$. A portfolio of cash and SPDR S\&P 500 ETF is investigated,
with annual risk free rate $r^{f}=0.012$, portfolio weight increment
$0.01$ and initial investment amount $W_{0^{-}}=\$100m$.}{\footnotesize \par}
\end{singlespace}
\end{figure}

Figure \ref{fig:liquidity-control} shows the time evolution of the
portfolio allocation distribution and the wealth distribution. When
the market liquidity effects are small ($\sigma_{\text{day}}=2.5,\text{Vol}_{\text{day}}=120m$,
left-hand side column), both the portfolio allocation and the wealth
are widely spread, along with large portfolio turnovers at the beginning
and at the end of the investment horizon. By contrast, for large market
liquidity effects ($\sigma_{\text{day}}=12.5,\text{Vol}_{\text{day}}=12m$,
right-hand side column) transactions become very costly and the algorithm
disallows large portfolio turnovers. As a consequence, the portfolio
allocation distribution is tightened at a relatively low level ($\alpha\simeq0.2$)
and its time evolution is smooth. Regarding the wealth distribution,
as expected, the less liquid the market, the lower the CER (as was
shown in Table \ref{tab:mc-liquidity}) and the lower the dispersion
of the wealth distribution. This is due to the lower and more stable
allocation in stock.

\subsection{Certainty equivalent losses associated with ignoring liquidity effects\label{subsec:CER-losses}}

Table \ref{tab:utility-loss} compares the CER of an investor who
takes heed of liquidity effects when making allocation decisions to
the CER of an investor who ignores liquidity effects. For the investor
who ignores liquidity effects, we set $\text{LC}=\text{MI}=0$ in
the LSMC algorithm, then reset $\text{LC}=\text{LC}\left(\Delta q\right)$
and $\text{MI}=\text{MI}\left(\Delta q\right)$ for calculating CER.
Unsurprisingly, the liquidity-aware investor always has a positive
CER, while the CER of the liquidity-blind investor can reach negative
territory in illiquid markets. The massive gain in CER of liquidity-aware
portfolio allocation over liquidity-blind portfolio allocation illustrates
how taking these costs into account is vital for reaching one's performance
target in real life situations where these liquidity effects do occur.
It also illustrates the ability of our algorithm to properly cope
with intermediate costs.

\subsection{Sensitivity analysis\label{subsec:Sensitivity}}

Table \ref{tab:sen-vol} reports the sensitivity of CER and of the
initial stock allocation $\alpha_{0}$ with respect to the daily trading
volume. The effect of increasing daily trading volume (and therefore
increasing liquidity, cf. equations \eqref{eq:impact-func}-\eqref{eq:liquidity-func})
on CER and $\alpha_{0}$ is consistent under different levels of daily
volatility: CER and $\alpha_{0}$ increase at diminishing rates. Similarly,
Table \ref{tab:sen-sig} shows that increasing daily volatility (and
therefore decreasing liquidity, cf. equations \eqref{eq:impact-func}-\eqref{eq:liquidity-func})
decreases CER and $\alpha_{0}$ at diminishing rates. 

Table \ref{tab:sen-W} reports the sensitivity of CER and the initial
stock allocation $\alpha_{0}$ with respect to the initial investment
amount $W_{0^{-}}$. As expected, the CER and $\alpha_{0}$ decrease
with respect to $W_{0^{-}}$ due to the larger liquidity effects on
bigger portfolios. Under extreme liquidity effects $(\sigma_{\text{day}},\text{Vol}_{\text{day}})=\left(12.5,12m\right)$,
$\alpha_{0}$ remains zero for $W_{0^{-}}>600m$, meaning that the
quasi impossibility to rebalance the portfolio makes a full risk-free
allocation the best initial investment in terms of expected utility. 

Table \ref{tab:sen-T} reports the sensitivity of CER and $\alpha_{0}$
with respect to the investment horizon. The initial allocations quickly
converge to a certain level when the time horizon is increased: they
decrease towards this limit when liquidity effect is small ($\sigma_{\text{day}}=2.5,\text{Vol}_{\text{day}}=120m$)
while they increase towards this limit in the two other cases ($\sigma_{\text{day}}=7.5,\text{Vol}_{\text{day}}=55m$
and $\sigma_{\text{day}}=12.5,\text{Vol}_{\text{day}}=12m$). The
CER first increases then decreases with time horizon for the two cases
$\sigma_{\text{day}}=2.5,\text{Vol}_{\text{day}}=120m$ and $\sigma_{\text{day}}=7.5,\text{Vol}_{\text{day}}=55m$
while monotonically increases when liquidity effect is large ($\sigma_{\text{day}}=12.5,\text{Vol}_{\text{day}}=12m$). 

Finally, Table \ref{tab:sen-gamma} reports the sensitivity of CER
and $\alpha_{0}$ with respect to the risk-aversion level $\gamma$
of the CRRA utility. As expected, CER and $\alpha_{0}$ both decrease
under every liquidity situation when risk-aversion is increased.

To conclude this numerical section, we can emphasize that a key feature
of the general portfolio allocation algorithm proposed in this paper
is the ability to measure and account for the effect of imperfect
liquidity on dynamic portfolio allocation. After having validated
the stability and convergence of the algorithm, we were able to compute
and report the sensitivities of portfolio allocation and portfolio
performance with respect to various parameters. Such analyzes can
be adapted to different models, markets and investment styles, and
bring insights into the most advantageous way to adjust dynamic portfolio
allocation in less liquid markets.

\section{Conclusion\label{sec:Conclusion}}

This paper describes a simulation-and-regression method for solving
portfolio allocation problems with general transaction costs, temporary
liquidity costs and permanent market impacts. To deal with permanent
market impacts, we model the price dynamics as endogenous state variables
which are separate from the exogenous return dynamics, while maintain
the same computational complexity of the algorithm with these additional
endogenous variables. The simulation nature of the chosen algorithm
makes it suitable for multivariate portfolios with realistic asset
dynamics and realistic liquidity effects. The algorithm adapts \citet{Kharroubi2014}'s
control randomization approach to the discrete portfolio allocation.
For each allocation level, the endogenous state variables are correspondingly
updated and are used to estimate the value function by a simple linear
least-squares regression. We iterate the whole algorithm by using
the optimal control estimates of the first run as the initial controls
of the second run. Our numerical tests show that, with second-order
polynomial basis, the proposed control discretization combined with
global control iteration outperforms the control regression approach
of \citet{Kharroubi2014}, all the more so with highly nonlinear utility
functions (high risk-aversion). 

We apply our method to solve a realistic cash-and-stock portfolio
with the power-law liquidity model of \citet{Almgren2005}. We show
that the losses associated with ignoring liquidity effects can be
substantial, indicating the necessity to account for liquidity effects
when making portfolio allocation decisions in real markets. Most importantly,
our algorithm is able to protect the portfolio value in illiquid markets.
Going further, we analyze the sensitivities of certainty equivalent
returns and optimal allocations with respect to trading volume, stock
price volatility, initial capital, risk-aversion level and investment
horizon.

The flexibility of the algorithm motivates future studies to investigate
alternative portfolio performance measures beyond expected utility,
alternative liquidity models, or to incorporate additional features
such as cross-asset price impact. It could also be easily adapted
to the problems of optimal portfolio liquidation and more general
optimal switching problems with endogenous uncertainty.

\newpage

\bibliographystyle{chicago}
\bibliography{bib_portfolio}

\newpage

\begin{table}[H]
\caption{\label{tab:assets}Return predictors (exogenous state variables)}

\smallskip
\begin{singlespace}
\centering{}%
\begin{tabular}{lllll}
\multirow{2}{*}{{\footnotesize{}Return Predictors}} &  & \multirow{2}{*}{{\footnotesize{}ETF Name}} &  & \multirow{2}{*}{{\footnotesize{}ETF Ticker}}\tabularnewline
 &  &  &  & \tabularnewline
\hline 
{\footnotesize{}U.S. stock} &  & {\footnotesize{}SPDR S\&P 500 ETF} &  & {\footnotesize{}SPY}\tabularnewline
{\footnotesize{}U.S. bond} &  & {\footnotesize{}Vanguard Total Bond Market ETF} &  & {\footnotesize{}BND}\tabularnewline
{\footnotesize{}International stock} &  & {\footnotesize{}iShares MSCI EAFE ETF} &  & {\footnotesize{}EFA}\tabularnewline
{\footnotesize{}Emerging market stock} &  & {\footnotesize{}iShares MSCI Emerging Markets ETF} &  & {\footnotesize{}EEM}\tabularnewline
{\footnotesize{}Gold} &  & {\footnotesize{}SPDR Gold Shares ETF} &  & {\footnotesize{}GLD}\tabularnewline
{\footnotesize{}International bond} &  & {\footnotesize{}SPDR Barclays International Treasury Bond ETF} &  & {\footnotesize{}BWX}\tabularnewline
{\footnotesize{}Silver} &  & {\footnotesize{}iShares Silver Trust ETF} &  & {\footnotesize{}SLV}\tabularnewline
{\footnotesize{}Crude oil} &  & {\footnotesize{}U.S. Oil ETF} &  & {\footnotesize{}USO}\tabularnewline
{\footnotesize{}U.S. dollar} &  & {\footnotesize{}PowerShares Deutsche Bank U.S. Dollar Bullish ETF} &  & {\footnotesize{}UUP}\tabularnewline
{\footnotesize{}Euro} &  & {\footnotesize{}CurrencyShares Euro ETF} &  & {\footnotesize{}FXE}\tabularnewline
{\footnotesize{}Japanese Yen} &  & {\footnotesize{}CurrencyShares Japanese Yen ETF} &  & {\footnotesize{}FXY}\tabularnewline
{\footnotesize{}Australian dollar} &  & {\footnotesize{}CurrencyShares Australian dollar ETF} &  & {\footnotesize{}FXA}\tabularnewline
\hline 
\end{tabular}
\end{singlespace}
\end{table}

\begin{table}[H]
\begin{centering}
\caption{\label{tab:mc-benchmark}Monte Carlo convergence with respect to risk
aversion level and time horizon}
\begin{tabular}{cccrrrrrrrrrrr}
 &  &  & {\footnotesize{}$\gamma=5$} &  &  &  &  &  & {\footnotesize{}$\gamma=15$} &  &  &  & \tabularnewline
\hline 
\multirow{2}{*}{} & \multirow{2}{*}{} & \multirow{2}{*}{} & \multirow{2}{*}{{\footnotesize{}KLP}} & \multirow{2}{*}{{\footnotesize{}$I=0$}} & \multirow{2}{*}{{\footnotesize{}$I=1$}} & \multirow{2}{*}{{\footnotesize{}$I=2$}} & \multirow{2}{*}{{\footnotesize{}$I=3$}} & \multirow{2}{*}{} & \multirow{2}{*}{{\footnotesize{}KLP}} & \multirow{2}{*}{{\footnotesize{}$I=0$}} & \multirow{2}{*}{{\footnotesize{}$I=1$}} & \multirow{2}{*}{{\footnotesize{}$I=2$}} & \multirow{2}{*}{{\footnotesize{}$I=3$}}\tabularnewline
 &  &  &  &  &  &  &  &  &  &  &  &  & \tabularnewline
\hline 
\multirow{5}{*}{{\footnotesize{}$N=5$}} & {\footnotesize{}M=$10^{3}$} &  & {\footnotesize{}1.45} & {\footnotesize{}1.54} & {\footnotesize{}1.55} & {\footnotesize{}1.58} & {\footnotesize{}1.58} &  & {\footnotesize{}1.02} & {\footnotesize{}1.19} & {\footnotesize{}1.25} & {\footnotesize{}1.31} & {\footnotesize{}1.34}\tabularnewline
 & {\footnotesize{}M=$10^{4}$} &  & {\footnotesize{}1.54} & {\footnotesize{}1.57} & {\footnotesize{}1.58} & {\footnotesize{}1.58} & {\footnotesize{}1.58} &  & {\footnotesize{}1.15} & {\footnotesize{}1.36} & {\footnotesize{}1.38} & {\footnotesize{}1.39} & {\footnotesize{}1.39}\tabularnewline
 & {\footnotesize{}M=$10^{5}$} &  & {\footnotesize{}1.54} & {\footnotesize{}1.58} & {\footnotesize{}1.58} & {\footnotesize{}1.58} & {\footnotesize{}1.58} &  & {\footnotesize{}1.18} & {\footnotesize{}1.38} & {\footnotesize{}1.39} & {\footnotesize{}1.39} & {\footnotesize{}1.39}\tabularnewline
 & {\footnotesize{}M=$10^{6}$} &  & {\footnotesize{}1.55} & {\footnotesize{}1.58} & {\footnotesize{}1.58} & {\footnotesize{}1.58} & {\footnotesize{}1.58} &  & {\footnotesize{}1.24} & {\footnotesize{}1.38} & {\footnotesize{}1.39} & {\footnotesize{}1.39} & {\footnotesize{}1.39}\tabularnewline
 &  &  &  &  &  & \multicolumn{2}{r}{{\footnotesize{}$\text{CER}^{*}=1.58$}} &  &  &  &  & \multicolumn{2}{r}{{\footnotesize{}$\text{CER}^{*}=1.39$}}\tabularnewline
\multirow{5}{*}{{\footnotesize{}$N=15$}} & {\footnotesize{}M=$10^{3}$} &  & {\footnotesize{}0.85} & {\footnotesize{}1.37} & {\footnotesize{}1.40} & {\footnotesize{}1.42} & {\footnotesize{}1.43} &  & {\footnotesize{}0.03} & {\footnotesize{}0.59} & {\footnotesize{}0.92} & {\footnotesize{}1.06} & {\footnotesize{}1.13}\tabularnewline
 & {\footnotesize{}M=$10^{4}$} &  & {\footnotesize{}1.33} & {\footnotesize{}1.49} & {\footnotesize{}1.51} & {\footnotesize{}1.52} & {\footnotesize{}1.53} &  & {\footnotesize{}0.45} & {\footnotesize{}1.01} & {\footnotesize{}1.13} & {\footnotesize{}1.20} & {\footnotesize{}1.25}\tabularnewline
 & {\footnotesize{}M=$10^{5}$} &  & {\footnotesize{}1.47} & {\footnotesize{}1.51} & {\footnotesize{}1.52} & {\footnotesize{}1.53} & {\footnotesize{}1.53} &  & {\footnotesize{}0.84} & {\footnotesize{}1.18} & {\footnotesize{}1.31} & {\footnotesize{}1.33} & {\footnotesize{}1.35}\tabularnewline
 & {\footnotesize{}M=$10^{6}$} &  & {\footnotesize{}1.48} & {\footnotesize{}1.52} & {\footnotesize{}1.53} & {\footnotesize{}1.53} & {\footnotesize{}1.53} &  & {\footnotesize{}1.04} & {\footnotesize{}1.32} & {\footnotesize{}1.36} & {\footnotesize{}1.37} & {\footnotesize{}1.37}\tabularnewline
 &  &  &  &  &  & \multicolumn{2}{r}{{\footnotesize{}$\text{CER}^{*}=1.53$}} &  &  &  &  & \multicolumn{2}{r}{{\footnotesize{}$\text{CER}^{*}=1.37$}}\tabularnewline
\hline 
\end{tabular}
\par\end{centering}
\smallskip

{\footnotesize{}This table compares the Monte Carlo convergence of
Algorithm \ref{algo:lsmc} with the control regression algorithm of
\citet{Kharroubi2014} (`KLP') with second-order polynomial basis.
$I=0,1,2,3$ denotes the number of additional control iterations,
and `CER{*}' stands for the best estimate. We assume a CARA utility
investor, i.e., $U(w)=-\exp(-\gamma w)$, a cash and stock portfolio
with no switching costs, risk-free rate $0.012$, and stock annual
return with mean $0.03$ and volatility $0.15$. The annually adjusted
certainty equivalent returns (in percentage points) are reported for
different Monte Carlo sample sizes ($M=10^{3},10^{4},10^{5},10^{6}$),
different investment horizons ($N=5\text{yrs},15\text{yrs}$) and
different risk-aversion parameters ($\gamma=5,15$). }{\footnotesize \par}
\end{table}

\newpage

\begin{landscape}

\begin{table}
{\small{}\caption{\label{tab:mc-liquidity}Monte Carlo convergence under liquidity cost
and market impact}
}{\small \par}
\begin{singlespace}
\begin{centering}
{\footnotesize{}}%
\begin{tabular}{ccccllllllllllllllll}
 &  &  & \multicolumn{5}{l}{{\footnotesize{}$\text{Vol}_{\text{day}}=120m$}} &  & \multicolumn{5}{l}{{\footnotesize{}$\text{Vol}_{\text{day}}=55m$}} &  & \multicolumn{5}{l}{{\footnotesize{}$\text{Vol}_{\text{day}}=12m$}}\tabularnewline
\hline 
\multirow{2}{*}{} & \multirow{2}{*}{} & \multirow{2}{*}{} & \multirow{2}{*}{{\footnotesize{}KLP}} & \multirow{2}{*}{{\footnotesize{}$I=0$}} & \multirow{2}{*}{{\footnotesize{}$I=1$}} & \multirow{2}{*}{{\footnotesize{}$I=2$}} & \multirow{2}{*}{{\footnotesize{}$I=3$}} & \multirow{2}{*}{} & \multirow{2}{*}{{\footnotesize{}KLP}} & \multirow{2}{*}{{\footnotesize{}$I=0$}} & \multirow{2}{*}{{\footnotesize{}$I=1$}} & \multirow{2}{*}{{\footnotesize{}$I=2$}} & \multirow{2}{*}{{\footnotesize{}$I=3$}} & \multirow{2}{*}{} & \multirow{2}{*}{{\footnotesize{}KLP}} & \multirow{2}{*}{{\footnotesize{}$I=0$}} & \multirow{2}{*}{{\footnotesize{}$I=1$}} & \multirow{2}{*}{{\footnotesize{}$I=2$}} & \multirow{2}{*}{{\footnotesize{}$I=3$}}\tabularnewline
 &  &  &  &  &  &  &  &  &  &  &  &  &  &  &  &  &  &  & \tabularnewline
\hline 
\multirow{4}{*}{{\footnotesize{}$\sigma_{\text{day}}=2.5$}} & {\footnotesize{}M=$10^{3}$} &  & {\footnotesize{}$33.3$} & {\footnotesize{}$55.0$} & {\footnotesize{}$56.9$} & {\footnotesize{}$57.7$} & {\footnotesize{}$57.8$} &  & {\footnotesize{}$31.7$} & {\footnotesize{}$56.9$} & {\footnotesize{}$58.7$} & {\footnotesize{}$58.7$} & {\footnotesize{}$58.5$} &  & {\footnotesize{}$21.9$} & {\footnotesize{}$41.6$} & {\footnotesize{}$42.6$} & {\footnotesize{}$43.6$} & {\footnotesize{}$43.6$}\tabularnewline
 & {\footnotesize{}M=$10^{4}$} &  & {\footnotesize{}$55.1$} & {\footnotesize{}$56.5$} & {\footnotesize{}$57.5$} & {\footnotesize{}$57.7$} & {\footnotesize{}$57.7$} &  & {\footnotesize{}$51.9$} & {\footnotesize{}$54.8$} & {\footnotesize{}$54.9$} & {\footnotesize{}$54.9$} & {\footnotesize{}$54.9$} &  & {\footnotesize{}$42.0$} & {\footnotesize{}$42.7$} & {\footnotesize{}$44.0$} & {\footnotesize{}$44.0$} & {\footnotesize{}$44.0$}\tabularnewline
 & {\footnotesize{}M=$10^{5}$} &  & {\footnotesize{}$57.0$} & {\footnotesize{}$57.5$} & {\footnotesize{}$57.8$} & {\footnotesize{}$57.8$} & {\footnotesize{}$57.8$} &  & {\footnotesize{}$53.5$} & {\footnotesize{}$54.1$} & {\footnotesize{}$54.2$} & {\footnotesize{}$54.2$} & {\footnotesize{}$54.2$} &  & {\footnotesize{}$42.7$} & {\footnotesize{}$43.5$} & {\footnotesize{}$44.2$} & {\footnotesize{}$44.2$} & {\footnotesize{}$44.2$}\tabularnewline
 & {\footnotesize{}M=$10^{6}$} &  & {\footnotesize{}$57.5$} & {\footnotesize{}$57.5$} & {\footnotesize{}$57.9$} & {\footnotesize{}$57.9$} & {\footnotesize{}$57.9$} &  & {\footnotesize{}$54.1$} & {\footnotesize{}$54.2$} & {\footnotesize{}$54.5$} & {\footnotesize{}$54.5$} & {\footnotesize{}$54.5$} &  & {\footnotesize{}$43.7$} & {\footnotesize{}$44.1$} & {\footnotesize{}$44.5$} & {\footnotesize{}$44.5$} & {\footnotesize{}$44.5$}\tabularnewline
 &  &  &  &  &  &  &  &  &  &  &  &  &  &  &  &  &  &  & \tabularnewline
\multirow{4}{*}{{\footnotesize{}$\sigma_{\text{day}}=7.5$}} & {\footnotesize{}M=$10^{3}$} &  & {\footnotesize{}$21.3$} & {\footnotesize{}$47.0$} & {\footnotesize{}$47.6$} & {\footnotesize{}$47.6$} & {\footnotesize{}$47.6$} &  & {\footnotesize{}$20.3$} & {\footnotesize{}$39.7$} & {\footnotesize{}$41.5$} & {\footnotesize{}$41.5$} & {\footnotesize{}$41.5$} &  & {\footnotesize{}$17.9$} & {\footnotesize{}$28.2$} & {\footnotesize{}$29.4$} & {\footnotesize{}$29.4$} & {\footnotesize{}$29.5$}\tabularnewline
 & {\footnotesize{}M=$10^{4}$} &  & {\footnotesize{}$45.4$} & {\footnotesize{}$47.2$} & {\footnotesize{}$47.6$} & {\footnotesize{}$47.6$} & {\footnotesize{}$47.6$} &  & {\footnotesize{}$39.7$} & {\footnotesize{}$40.2$} & {\footnotesize{}$41.7$} & {\footnotesize{}$41.7$} & {\footnotesize{}$41.7$} &  & {\footnotesize{}$28.1$} & {\footnotesize{}$28.6$} & {\footnotesize{}$29.6$} & {\footnotesize{}$29.6$} & {\footnotesize{}$29.6$}\tabularnewline
 & {\footnotesize{}M=$10^{5}$} &  & {\footnotesize{}$47.1$} & {\footnotesize{}$47.4$} & {\footnotesize{}$47.6$} & {\footnotesize{}$47.6$} & {\footnotesize{}$47.6$} &  & {\footnotesize{}$40.4$} & {\footnotesize{}$41.0$} & {\footnotesize{}$41.8$} & {\footnotesize{}$41.8$} & {\footnotesize{}$41.8$} &  & {\footnotesize{}$28.6$} & {\footnotesize{}$29.0$} & {\footnotesize{}$29.9$} & {\footnotesize{}$29.9$} & {\footnotesize{}$29.9$}\tabularnewline
 & {\footnotesize{}M=$10^{6}$} &  & {\footnotesize{}$47.4$} & {\footnotesize{}$47.5$} & {\footnotesize{}$47.9$} & {\footnotesize{}$47.9$} & {\footnotesize{}$47.9$} &  & {\footnotesize{}$41.0$} & {\footnotesize{}$41.6$} & {\footnotesize{}$42.0$} & {\footnotesize{}$42.0$} & {\footnotesize{}$42.0$} &  & {\footnotesize{}$29.0$} & {\footnotesize{}$29.2$} & {\footnotesize{}$30.0$} & {\footnotesize{}$30.0$} & {\footnotesize{}$30.0$}\tabularnewline
 &  &  &  &  &  &  &  &  &  &  &  &  &  &  &  &  &  &  & \tabularnewline
\multirow{4}{*}{{\footnotesize{}$\sigma_{\text{day}}=12.5$}} & {\footnotesize{}M=$10^{3}$} &  & {\footnotesize{}$20.0$} & {\footnotesize{}$39.9$} & {\footnotesize{}$40.4$} & {\footnotesize{}$40.9$} & {\footnotesize{}$41.2$} &  & {\footnotesize{}$20.4$} & {\footnotesize{}$31.0$} & {\footnotesize{}$33.5$} & {\footnotesize{}$33.7$} & {\footnotesize{}$34.1$} &  & {\footnotesize{}$12.8$} & {\footnotesize{}$19.0$} & {\footnotesize{}$23.5$} & {\footnotesize{}$23.7$} & {\footnotesize{}$23.7$}\tabularnewline
 & {\footnotesize{}M=$10^{4}$} &  & {\footnotesize{}$39.2$} & {\footnotesize{}$40.4$} & {\footnotesize{}$41.1$} & {\footnotesize{}$41.1$} & {\footnotesize{}$41.2$} &  & {\footnotesize{}$33.0$} & {\footnotesize{}$32.2$} & {\footnotesize{}$33.9$} & {\footnotesize{}$34.4$} & {\footnotesize{}$35.1$} &  & {\footnotesize{}$20.3$} & {\footnotesize{}$20.7$} & {\footnotesize{}$23.6$} & {\footnotesize{}$23.7$} & {\footnotesize{}$23.8$}\tabularnewline
 & {\footnotesize{}M=$10^{5}$} &  & {\footnotesize{}$40.5$} & {\footnotesize{}$40.8$} & {\footnotesize{}$41.2$} & {\footnotesize{}$41.2$} & {\footnotesize{}$41.2$} &  & {\footnotesize{}$33.5$} & {\footnotesize{}$33.6$} & {\footnotesize{}$35.1$} & {\footnotesize{}$35.1$} & {\footnotesize{}$35.1$} &  & {\footnotesize{}$21.6$} & {\footnotesize{}$21.9$} & {\footnotesize{}$24.0$} & {\footnotesize{}$24.1$} & {\footnotesize{}$24.1$}\tabularnewline
 & {\footnotesize{}M=$10^{6}$} &  & {\footnotesize{}$41.0$} & {\footnotesize{}$41.1$} & {\footnotesize{}$41.3$} & {\footnotesize{}$41.3$} & {\footnotesize{}$41.3$} &  & {\footnotesize{}$33.9$} & {\footnotesize{}$33.9$} & {\footnotesize{}$35.1$} & {\footnotesize{}$35.2$} & {\footnotesize{}$35.2$} &  & {\footnotesize{}$22.0$} & {\footnotesize{}$22.1$} & {\footnotesize{}$24.1$} & {\footnotesize{}$24.2$} & {\footnotesize{}$24.2$}\tabularnewline
\hline 
\end{tabular}
\par\end{centering}{\footnotesize \par}
\end{singlespace}
\smallskip

\begin{singlespace}
{\footnotesize{}This table compares the Monte Carlo convergence of
the monthly adjusted certainty equivalent return (in basis points)
for the CRRA utility with $\gamma=5$ and investment horizon $N=12$
months, under different daily volatilities $(\sigma_{\text{day}}=2.5,7.5,12.5)$
and different daily trading volumes $(\text{Vol}_{\text{day}}=120m,55m,12m)$,
where the Monte Carlo procedure is performed under different sizes
of Monte Carlo sample ($M=10^{3},10^{4},10^{5},10^{6}$) and different
iterations $(I=0,1,2,3)$. A portfolio of cash and SPDR S\&P 500 ETF
is investigated, with annual risk free rate $r^{f}=0.012$, portfolio
weight increment $0.01$ and initial investment amount $W_{0^{-}}=\$100m$.}{\footnotesize \par}
\end{singlespace}
\end{table}

\end{landscape}

\newpage

\begin{landscape}

\begin{table}
{\small{}\caption{\label{tab:utility-loss}Certainty equivalent losses with ignoring
liquidity effects}
}{\small \par}

\smallskip
\begin{singlespace}
\begin{centering}
{\footnotesize{}}%
\begin{tabular}{cc>{\centering}p{0.05cm}>{\raggedleft}p{0.9cm}>{\raggedleft}p{0.75cm}>{\raggedleft}p{0.75cm}>{\raggedleft}p{1.1cm}>{\raggedleft}p{0.75cm}>{\raggedleft}p{0.75cm}>{\raggedleft}p{0.05cm}>{\raggedleft}p{0.9cm}>{\raggedleft}p{0.75cm}>{\raggedleft}p{0.75cm}>{\raggedleft}p{1.1cm}>{\raggedleft}p{0.75cm}>{\raggedleft}p{0.75cm}}
 &  &  & \multicolumn{6}{l}{{\footnotesize{}\hspace{1.25em}$\gamma=5$}} &  & \multicolumn{6}{l}{{\footnotesize{}\hspace{1.25em}$\gamma=10$}}\tabularnewline
\hline 
 &  &  & \multicolumn{3}{l}{{\footnotesize{}\hspace{1.25em}Liquidity-aware}} & \multicolumn{3}{l}{{\footnotesize{}\hspace{1.9em}Liquidity-blind}} &  & \multicolumn{3}{l}{{\footnotesize{}\hspace{1.25em}Liquidity-aware}} & \multicolumn{3}{l}{{\footnotesize{}\hspace{1.9em}Liquidity-blind}}\tabularnewline
\hline 
\multirow{2}{*}{{\footnotesize{}$\text{Vol}_{\text{day}}$}} & \multirow{2}{*}{{\footnotesize{}$W_{0^{-}}$}} &  & {\footnotesize{}$\sigma_{\text{day}}$} & {\footnotesize{}$\sigma_{\text{day}}$} & {\footnotesize{}$\sigma_{\text{day}}$} & {\footnotesize{}$\sigma_{\text{day}}$} & {\footnotesize{}$\sigma_{\text{day}}$} & {\footnotesize{}$\sigma_{\text{day}}$} &  & {\footnotesize{}$\sigma_{\text{day}}$} & {\footnotesize{}$\sigma_{\text{day}}$} & {\footnotesize{}$\sigma_{\text{day}}$} & {\footnotesize{}$\sigma_{\text{day}}$} & {\footnotesize{}$\sigma_{\text{day}}$} & {\footnotesize{}$\sigma_{\text{day}}$}\tabularnewline
 &  &  & {\footnotesize{}=2.5} & {\footnotesize{}=7.5} & {\footnotesize{}=12.5} & {\footnotesize{}=2.5} & {\footnotesize{}=7.5} & {\footnotesize{}=12.5} &  & {\footnotesize{}=2.5} & {\footnotesize{}=7.5} & {\footnotesize{}=12.5} & {\footnotesize{}=2.5} & {\footnotesize{}=7.5} & {\footnotesize{}=12.5}\tabularnewline
\hline 
 & {\footnotesize{}0.1b} &  & {\footnotesize{}$57.8$} & {\footnotesize{}$47.6$} & {\footnotesize{}$41.2$} & {\footnotesize{}$56.6$} & {\footnotesize{}$38.5$} & {\footnotesize{}$21.2$} &  & {\footnotesize{}$37.7$} & {\footnotesize{}$31.6$} & {\footnotesize{}$27.8$} & {\footnotesize{}$36.2$} & {\footnotesize{}$25.4$} & {\footnotesize{}$15.3$}\tabularnewline
{\footnotesize{}$120m$} & {\footnotesize{}0.5b} &  & {\footnotesize{}$49.2$} & {\footnotesize{}$35.3$} & {\footnotesize{}$28.5$} & {\footnotesize{}$41.8$} & {\footnotesize{}$-2.4$} & {\footnotesize{}$-41.0$} &  & {\footnotesize{}$32.6$} & {\footnotesize{}$24.2$} & {\footnotesize{}$20.1$} & {\footnotesize{}$27.4$} & {\footnotesize{}$1.7$} & {\footnotesize{}$-19.9$}\tabularnewline
 & {\footnotesize{}1.0b} &  & {\footnotesize{}$44.2$} & {\footnotesize{}$29.9$} & {\footnotesize{}$24.0$} & {\footnotesize{}$29.9$} & {\footnotesize{}$-32.7$} & {\footnotesize{}$-82.6$} &  & {\footnotesize{}$29.6$} & {\footnotesize{}$20.9$} & {\footnotesize{}$16.4$} & {\footnotesize{}$20.4$} & {\footnotesize{}$-15.3$} & {\footnotesize{}$-42.9$}\tabularnewline
 &  &  &  &  &  &  &  &  &  &  &  &  &  &  & \tabularnewline
 & {\footnotesize{}0.1b} &  & {\footnotesize{}$54.2$} & {\footnotesize{}$41.8$} & {\footnotesize{}$35.1$} & {\footnotesize{}$51.1$} & {\footnotesize{}$23.0$} & {\footnotesize{}$-3.2$} &  & {\footnotesize{}$35.5$} & {\footnotesize{}$28.2$} & {\footnotesize{}$24.1$} & {\footnotesize{}$32.9$} & {\footnotesize{}$16.3$} & {\footnotesize{}$1.2$}\tabularnewline
{\footnotesize{}$55m$} & {\footnotesize{}0.5b} &  & {\footnotesize{}$43.5$} & {\footnotesize{}$29.3$} & {\footnotesize{}$23.5$} & {\footnotesize{}$28.1$} & {\footnotesize{}$-37.2$} & {\footnotesize{}$-88.4$} &  & {\footnotesize{}$29.2$} & {\footnotesize{}$20.4$} & {\footnotesize{}$16.0$} & {\footnotesize{}$19.3$} & {\footnotesize{}$-17.8$} & {\footnotesize{}$-46.1$}\tabularnewline
 & {\footnotesize{}1.0b} &  & {\footnotesize{}$38.1$} & {\footnotesize{}$24.5$} & {\footnotesize{}$19.5$} & {\footnotesize{}$-2.0$} & {\footnotesize{}$-77.9$} & {\footnotesize{}$-137.3$} &  & {\footnotesize{}$25.9$} & {\footnotesize{}$16.8$} & {\footnotesize{}$5.7$} & {\footnotesize{}$8.7$} & {\footnotesize{}$-40.3$} & {\footnotesize{}$-73.9$}\tabularnewline
 &  &  &  &  &  &  &  &  &  &  &  &  &  &  & \tabularnewline
 & {\footnotesize{}0.1b} &  & {\footnotesize{}$44.2$} & {\footnotesize{}$29.9$} & {\footnotesize{}$24.0$} & {\footnotesize{}$29.9$} & {\footnotesize{}$-32.7$} & {\footnotesize{}$-82.6$} &  & {\footnotesize{}$29.6$} & {\footnotesize{}$20.9$} & {\footnotesize{}$16.4$} & {\footnotesize{}$20.4$} & {\footnotesize{}$-15.3$} & {\footnotesize{}$-42.9$}\tabularnewline
{\footnotesize{}$12m$} & {\footnotesize{}0.5b} &  & {\footnotesize{}$31.6$} & {\footnotesize{}$19.6$} & {\footnotesize{}$14.1$} & {\footnotesize{}$-22.0$} & {\footnotesize{}$-135.6$} & {\footnotesize{}$-197.3$} &  & {\footnotesize{}$21.9$} & {\footnotesize{}$12.6$} & {\footnotesize{}$9.1$} & {\footnotesize{}$-9.3$} & {\footnotesize{}$-72.9$} & {\footnotesize{}$-112.1$}\tabularnewline
 & {\footnotesize{}1.0b} &  & {\footnotesize{}$26.5$} & {\footnotesize{}$15.6$} & {\footnotesize{}$7.3$} & {\footnotesize{}$-58.7$} & {\footnotesize{}$-186.0$} & {\footnotesize{}$-239.8$} &  & {\footnotesize{}$18.5$} & {\footnotesize{}$7.4$} & {\footnotesize{}$2.1$} & {\footnotesize{}$-29.7$} & {\footnotesize{}$-104.3$} & {\footnotesize{}$-150.9$}\tabularnewline
\hline 
\end{tabular}
\par\end{centering}{\footnotesize \par}
\end{singlespace}
\smallskip

\begin{singlespace}
{\footnotesize{}This table compares the monthly adjusted certainty
equivalent return (in basis points) for two CRRA investor with $\gamma=5,10$
and investment horizon $N=12$ months: the first one takes heed of
liquidity effects (liquidity-aware) while the second one ignores liquidity
effects (liquidity-blind). The results are compared under different
daily volatilities $(\sigma_{\text{day}}=2.5,7.5,12.5)$, different
daily trading volumes $(\text{Vol}_{\text{day}}=120m,55m,12m)$, and
different initial investment amount ($W_{0^{-}}=\$100m,500m,1b$),
using $M=10^{5}$ Monte Carlo simulations with one control iteration
$I=1$. A portfolio of cash and SPDR S\&P 500 ETF is investigated,
with annual risk free rate $r^{f}=0.012$, and portfolio weight increment
$0.01$.}{\footnotesize \par}
\end{singlespace}
\end{table}

\end{landscape}

\begin{table}[H]
{\small{}\caption{\label{tab:sen-vol}Sensitivity to daily trading volatility}
}{\small \par}

\smallskip
\begin{singlespace}
\begin{centering}
{\footnotesize{}}%
\begin{tabular}{cclllllll}
 &  & \multicolumn{3}{l}{{\footnotesize{}CER}} &  & \multicolumn{3}{l}{{\footnotesize{}Initial allocation $\alpha_{0}$}}\tabularnewline
\hline 
\multirow{2}{*}{{\footnotesize{}$\text{Vol}_{\text{day}}$}} &  & {\footnotesize{}$\sigma_{\text{day}}$} & {\footnotesize{}$\sigma_{\text{day}}$} & {\footnotesize{}$\sigma_{\text{day}}$} &  & {\footnotesize{}$\sigma_{\text{day}}$} & {\footnotesize{}$\sigma_{\text{day}}$} & {\footnotesize{}$\sigma_{\text{day}}$}\tabularnewline
 &  & {\footnotesize{}=2.5} & {\footnotesize{}=7.5} & {\footnotesize{}=12.5} &  & {\footnotesize{}=2.5} & {\footnotesize{}=7.5} & {\footnotesize{}=12.5}\tabularnewline
\hline 
{\footnotesize{}10m} &  & {\footnotesize{}$42.8$} & {\footnotesize{}$28.6$} & {\footnotesize{}$22.9$} &  & {\footnotesize{}$0.31$} & {\footnotesize{}$0.18$} & {\footnotesize{}$0.12$}\tabularnewline
{\footnotesize{}20m} &  & {\footnotesize{}$47.9$} & {\footnotesize{}$33.8$} & {\footnotesize{}$27.4$} &  & {\footnotesize{}$0.36$} & {\footnotesize{}$0.23$} & {\footnotesize{}$0.16$}\tabularnewline
{\footnotesize{}30m} &  & {\footnotesize{}$50.6$} & {\footnotesize{}$37.0$} & {\footnotesize{}$30.4$} &  & {\footnotesize{}$0.41$} & {\footnotesize{}$0.26$} & {\footnotesize{}$0.19$}\tabularnewline
{\footnotesize{}40m} &  & {\footnotesize{}$52.4$} & {\footnotesize{}$39.3$} & {\footnotesize{}$32.6$} &  & {\footnotesize{}$0.42$} & {\footnotesize{}$0.27$} & {\footnotesize{}$0.21$}\tabularnewline
{\footnotesize{}50m} &  & {\footnotesize{}$53.7$} & {\footnotesize{}$41.1$} & {\footnotesize{}$34.3$} &  & {\footnotesize{}$0.45$} & {\footnotesize{}$0.30$} & {\footnotesize{}$0.23$}\tabularnewline
{\footnotesize{}60m} &  & {\footnotesize{}$54.7$} & {\footnotesize{}$42.5$} & {\footnotesize{}$35.8$} &  & {\footnotesize{}$0.45$} & {\footnotesize{}$0.31$} & {\footnotesize{}$0.25$}\tabularnewline
{\footnotesize{}70m} &  & {\footnotesize{}$55.4$} & {\footnotesize{}$43.7$} & {\footnotesize{}$37.0$} &  & {\footnotesize{}$0.46$} & {\footnotesize{}$0.32$} & {\footnotesize{}$0.26$}\tabularnewline
{\footnotesize{}80m} &  & {\footnotesize{}$56.1$} & {\footnotesize{}$44.7$} & {\footnotesize{}$38.0$} &  & {\footnotesize{}$0.48$} & {\footnotesize{}$0.33$} & {\footnotesize{}$0.27$}\tabularnewline
{\footnotesize{}90m} &  & {\footnotesize{}$56.6$} & {\footnotesize{}$45.5$} & {\footnotesize{}$39.0$} &  & {\footnotesize{}$0.50$} & {\footnotesize{}$0.34$} & {\footnotesize{}$0.27$}\tabularnewline
{\footnotesize{}100m} &  & {\footnotesize{}$57.1$} & {\footnotesize{}$46.3$} & {\footnotesize{}$39.0$} &  & {\footnotesize{}$0.51$} & {\footnotesize{}$0.34$} & {\footnotesize{}$0.28$}\tabularnewline
{\footnotesize{}110m} &  & {\footnotesize{}$57.5$} & {\footnotesize{}$47.0$} & {\footnotesize{}$40.6$} &  & {\footnotesize{}$0.52$} & {\footnotesize{}$0.35$} & {\footnotesize{}$0.29$}\tabularnewline
{\footnotesize{}120m} &  & {\footnotesize{}$57.8$} & {\footnotesize{}$47.6$} & {\footnotesize{}$41.1$} &  & {\footnotesize{}$0.53$} & {\footnotesize{}$0.37$} & {\footnotesize{}$0.29$}\tabularnewline
\hline 
\end{tabular}
\par\end{centering}{\footnotesize \par}
\end{singlespace}
\smallskip

\begin{singlespace}
{\footnotesize{}This table reports the sensitivity of the monthly
adjusted certainty equivalent return (in basis points) and the initial
stock allocation with respect to the daily trading volume $\text{Vol}_{\text{day}}=10m,20m,...,120m$,
for a CRRA utility with $\gamma=5$ and investment horizon $N=12$
months under different daily volatilities $(\sigma_{\text{day}}=2.5,7.5,12.5)$,
using $M=10^{5}$ Monte Carlo simulations with one control iteration
$I=1$. A portfolio of cash and SPDR S\&P 500 ETF is investigated,
with annual risk free rate $r^{f}=0.012$, portfolio weight increment
$0.01$ and initial investment amount $W_{0^{-}}=\$100m$.}{\footnotesize \par}
\end{singlespace}
\end{table}
\begin{table}[H]
{\small{}\caption{\label{tab:sen-sig}Sensitivity to daily volatility}
}{\small \par}

\smallskip
\begin{singlespace}
\begin{centering}
{\footnotesize{}}%
\begin{tabular}{cclllllll}
 &  & \multicolumn{3}{l}{{\footnotesize{}CER}} &  & \multicolumn{3}{l}{{\footnotesize{}Initial allocation $\alpha_{0}$}}\tabularnewline
\hline 
\multirow{2}{*}{{\footnotesize{}$\sigma_{\text{day}}$}} &  & {\footnotesize{}$\text{Vol}_{\text{day}}$} & {\footnotesize{}$\text{Vol}_{\text{day}}$} & {\footnotesize{}$\text{Vol}_{\text{day}}$} &  & {\footnotesize{}$\text{Vol}_{\text{day}}$} & {\footnotesize{}$\text{Vol}_{\text{day}}$} & {\footnotesize{}$\text{Vol}_{\text{day}}$}\tabularnewline
 &  & {\footnotesize{}=120m} & {\footnotesize{}=55m} & {\footnotesize{}=12m} &  & {\footnotesize{}=120m} & {\footnotesize{}=55m} & {\footnotesize{}=12m}\tabularnewline
\hline 
{\footnotesize{}2} &  & {\footnotesize{}$59.2$} & {\footnotesize{}$56.1$} & {\footnotesize{}$46.9$} &  & {\footnotesize{}$0.55$} & {\footnotesize{}$0.48$} & {\footnotesize{}$0.35$}\tabularnewline
{\footnotesize{}3} &  & {\footnotesize{}$56.5$} & {\footnotesize{}$52.5$} & {\footnotesize{}$41.8$} &  & {\footnotesize{}$0.50$} & {\footnotesize{}$0.44$} & {\footnotesize{}$0.30$}\tabularnewline
{\footnotesize{}4} &  & {\footnotesize{}$54.2$} & {\footnotesize{}$49.5$} & {\footnotesize{}$38.1$} &  & {\footnotesize{}$0.45$} & {\footnotesize{}$0.39$} & {\footnotesize{}$0.25$}\tabularnewline
{\footnotesize{}5} &  & {\footnotesize{}$52.1$} & {\footnotesize{}$46.9$} & {\footnotesize{}$35.1$} &  & {\footnotesize{}$0.42$} & {\footnotesize{}$0.36$} & {\footnotesize{}$0.24$}\tabularnewline
{\footnotesize{}6} &  & {\footnotesize{}$50.2.$} & {\footnotesize{}$44.6$} & {\footnotesize{}$32.7$} &  & {\footnotesize{}$0.40$} & {\footnotesize{}$0.33$} & {\footnotesize{}$0.22$}\tabularnewline
{\footnotesize{}7} &  & {\footnotesize{}$48.4$} & {\footnotesize{}$42.7$} & {\footnotesize{}$30.8$} &  & {\footnotesize{}$0.39$} & {\footnotesize{}$0.31$} & {\footnotesize{}$0.20$}\tabularnewline
{\footnotesize{}8} &  & {\footnotesize{}$46.9$} & {\footnotesize{}$41.0$} & {\footnotesize{}$29.1$} &  & {\footnotesize{}$0.35$} & {\footnotesize{}$0.29$} & {\footnotesize{}$0.18$}\tabularnewline
{\footnotesize{}9} &  & {\footnotesize{}$45.4$} & {\footnotesize{}$39.4$} & {\footnotesize{}$27.7$} &  & {\footnotesize{}$0.34$} & {\footnotesize{}$0.28$} & {\footnotesize{}$0.17$}\tabularnewline
{\footnotesize{}10} &  & {\footnotesize{}$44.1$} & {\footnotesize{}$38.0$} & {\footnotesize{}$26.5$} &  & {\footnotesize{}$0.33$} & {\footnotesize{}$0.27$} & {\footnotesize{}$0.16$}\tabularnewline
{\footnotesize{}11} &  & {\footnotesize{}$42.9$} & {\footnotesize{}$36.7$} & {\footnotesize{}$25.4$} &  & {\footnotesize{}$0.31$} & {\footnotesize{}$0.26$} & {\footnotesize{}$0.14$}\tabularnewline
{\footnotesize{}12} &  & {\footnotesize{}$41.8$} & {\footnotesize{}$35.6$} & {\footnotesize{}$24.5$} &  & {\footnotesize{}$0.31$} & {\footnotesize{}$0.25$} & {\footnotesize{}$0.14$}\tabularnewline
{\footnotesize{}13} &  & {\footnotesize{}$40.7$} & {\footnotesize{}$34.6$} & {\footnotesize{}$23.6$} &  & {\footnotesize{}$0.30$} & {\footnotesize{}$0.23$} & {\footnotesize{}$0.12$}\tabularnewline
\hline 
\end{tabular}
\par\end{centering}{\footnotesize \par}
\end{singlespace}
\smallskip

\begin{singlespace}
{\footnotesize{}This table reports the sensitivity of the monthly
adjusted certainty equivalent return (in basis points) and the initial
stock allocation with respect to the daily volatility $\sigma_{\text{day}}=2,3,...,13$,
for a CRRA utility with $\gamma=5$ and investment horizon$N=12$
months under different daily trading volume $(\text{Vol}_{\text{day}}=120m,55m,12m)$,
using $M=10^{5}$ Monte Carlo simulations with one control iteration
$I=1$. A portfolio of cash and SPDR S\&P 500 ETF is investigated,
with annual risk free rate $r^{f}=0.012$, portfolio weight increment
$0.01$ and initial investment amount $W_{0^{-}}=\$100m$.}{\footnotesize \par}
\end{singlespace}
\end{table}
\begin{table}[H]
{\small{}\caption{\label{tab:sen-W}Sensitivity to Investment Amount}
}{\small \par}

\smallskip
\begin{singlespace}
\begin{centering}
{\footnotesize{}}%
\begin{tabular}{ccllrllll}
 &  & \multicolumn{3}{l}{{\footnotesize{}CER}} &  & \multicolumn{3}{l}{{\footnotesize{}Initial allocation $\alpha_{0}$}}\tabularnewline
\hline 
\multirow{2}{*}{{\footnotesize{}$W_{0^{-}}$}} & {\footnotesize{}$\sigma_{\text{day}}$} & {\footnotesize{}$2.5$} & {\footnotesize{}$7.5$} & {\footnotesize{}$12.5$} &  & {\footnotesize{}$2.5$} & {\footnotesize{}$7.5$} & {\footnotesize{}$12.5$}\tabularnewline
 & {\footnotesize{}$\text{Vol}_{\text{day}}$} & {\footnotesize{}$120m$} & {\footnotesize{}$55m$} & {\footnotesize{}$12m$} &  & {\footnotesize{}$120m$} & {\footnotesize{}$55m$} & {\footnotesize{}$12m$}\tabularnewline
\hline 
{\footnotesize{}\$100m} &  & {\footnotesize{}$57.8$} & {\footnotesize{}$41.8$} & {\footnotesize{}$24.0$} &  & {\footnotesize{}$0.52$} & {\footnotesize{}$0.31$} & {\footnotesize{}$0.13$}\tabularnewline
{\footnotesize{}\$200m} &  & {\footnotesize{}$54.7$} & {\footnotesize{}$36.3$} & {\footnotesize{}$19.9$} &  & {\footnotesize{}$0.47$} & {\footnotesize{}$0.25$} & {\footnotesize{}$0.09$}\tabularnewline
{\footnotesize{}\$300m} &  & {\footnotesize{}$52.4$} & {\footnotesize{}$33.1$} & {\footnotesize{}$17.8$} &  & {\footnotesize{}$0.43$} & {\footnotesize{}$0.22$} & {\footnotesize{}$0.07$}\tabularnewline
{\footnotesize{}\$400m} &  & {\footnotesize{}$50.6$} & {\footnotesize{}$30.9$} & {\footnotesize{}$16.1$} &  & {\footnotesize{}$0.40$} & {\footnotesize{}$0.20$} & {\footnotesize{}$0.04$}\tabularnewline
{\footnotesize{}\$500m} &  & {\footnotesize{}$49.2$} & {\footnotesize{}$29.3$} & {\footnotesize{}$14.1$} &  & {\footnotesize{}$0.38$} & {\footnotesize{}$0.19$} & {\footnotesize{}$0.02$}\tabularnewline
{\footnotesize{}\$600m} &  & {\footnotesize{}$47.9$} & {\footnotesize{}$28.0$} & {\footnotesize{}$12.0$} &  & {\footnotesize{}$0.36$} & {\footnotesize{}$0.18$} & {\footnotesize{}$0.01$}\tabularnewline
{\footnotesize{}\$700m} &  & {\footnotesize{}$46.8$} & {\footnotesize{}$26.9$} & {\footnotesize{}$10.8$} &  & {\footnotesize{}$0.35$} & {\footnotesize{}$0.16$} & {\footnotesize{}$0.00$}\tabularnewline
{\footnotesize{}\$800m} &  & {\footnotesize{}$45.8$} & {\footnotesize{}$25.9$} & {\footnotesize{}$10.3$} &  & {\footnotesize{}$0.34$} & {\footnotesize{}$0.15$} & {\footnotesize{}$0.00$}\tabularnewline
{\footnotesize{}\$900m} &  & {\footnotesize{}$45.0$} & {\footnotesize{}$25.2$} & {\footnotesize{}$8.3$} &  & {\footnotesize{}$0.33$} & {\footnotesize{}$0.14$} & {\footnotesize{}$0.00$}\tabularnewline
{\footnotesize{}\$1b} &  & {\footnotesize{}$44.2$} & {\footnotesize{}$24.5$} & {\footnotesize{}$7.3$} &  & {\footnotesize{}$0.32$} & {\footnotesize{}$0.13$} & {\footnotesize{}$0.00$}\tabularnewline
\hline 
\end{tabular}
\par\end{centering}{\footnotesize \par}
\end{singlespace}
\smallskip

\begin{singlespace}
{\footnotesize{}This table reports the sensitivity of the monthly
adjusted certainty equivalent return (in basis points) and the initial
stock allocation with respect to the initial investment amount $W_{0^{-}}=\$100,200,...,1000m$,
for a CRRA utility with $\gamma=5$ and investment horizon$N=12$
months under different liquidity settings $(\sigma_{\text{day}},\text{Vol}_{\text{day}})=\left(2.5,120m\right),(7.5,55m),(12.5,12m)$,
using $M=10^{5}$ Monte Carlo simulations with one control iteration
$I=1$. A portfolio of cash and SPDR S\&P 500 ETF is investigated,
with annual risk free rate $r^{f}=0.012$, portfolio weight increment
$0.01$.}{\footnotesize \par}
\end{singlespace}
\end{table}
 
\begin{table}[H]
{\small{}\caption{\label{tab:sen-T}Sensitivity to investment horizon}
}{\small \par}

\smallskip
\begin{singlespace}
\begin{centering}
{\footnotesize{}}%
\begin{tabular}{cclllllll}
 &  & \multicolumn{3}{l}{{\footnotesize{}CER}} &  & \multicolumn{3}{l}{{\footnotesize{}Initial allocation $\alpha_{0}$}}\tabularnewline
\hline 
\multirow{2}{*}{$N$} & {\footnotesize{}$\sigma_{\text{day}}$} & {\footnotesize{}$2.5$} & {\footnotesize{}$7.5$} & {\footnotesize{}$12.5$} &  & {\footnotesize{}$2.5$} & {\footnotesize{}$7.5$} & {\footnotesize{}$12.5$}\tabularnewline
 & {\footnotesize{}$\text{Vol}_{\text{day}}$} & {\footnotesize{}$120m$} & {\footnotesize{}$55m$} & {\footnotesize{}$12m$} &  & {\footnotesize{}$120m$} & {\footnotesize{}$55m$} & {\footnotesize{}$12m$}\tabularnewline
\hline 
{\footnotesize{}2} &  & {\footnotesize{}$54.0$} & {\footnotesize{}$26.3$} & {\footnotesize{}$12.3$} &  & {\footnotesize{}$0.60$} & {\footnotesize{}$0.24$} & {\footnotesize{}$0.04$}\tabularnewline
{\footnotesize{}3} &  & {\footnotesize{}$67.0$} & {\footnotesize{}$39.8$} & {\footnotesize{}$16.0$} &  & {\footnotesize{}$0.58$} & {\footnotesize{}$0.30$} & {\footnotesize{}$0.07$}\tabularnewline
{\footnotesize{}4} &  & {\footnotesize{}$65.8$} & {\footnotesize{}$43.3$} & {\footnotesize{}$18.6$} &  & {\footnotesize{}$0.57$} & {\footnotesize{}$0.31$} & {\footnotesize{}$0.09$}\tabularnewline
{\footnotesize{}5} &  & {\footnotesize{}$63.7$} & {\footnotesize{}$43.8$} & {\footnotesize{}$20.4$} &  & {\footnotesize{}$0.55$} & {\footnotesize{}$0.31$} & {\footnotesize{}$0.11$}\tabularnewline
{\footnotesize{}6} &  & {\footnotesize{}$62.0$} & {\footnotesize{}$43.5$} & {\footnotesize{}$21.6$} &  & {\footnotesize{}$0.55$} & {\footnotesize{}$0.31$} & {\footnotesize{}$0.12$}\tabularnewline
{\footnotesize{}7} &  & {\footnotesize{}$60.8$} & {\footnotesize{}$43.3$} & {\footnotesize{}$22.4$} &  & {\footnotesize{}$0.54$} & {\footnotesize{}$0.31$} & {\footnotesize{}$0.13$}\tabularnewline
{\footnotesize{}8} &  & {\footnotesize{}$60.1$} & {\footnotesize{}$42.9$} & {\footnotesize{}$23.0$} &  & {\footnotesize{}$0.54$} & {\footnotesize{}$0.31$} & {\footnotesize{}$0.13$}\tabularnewline
{\footnotesize{}9} &  & {\footnotesize{}$59.4$} & {\footnotesize{}$42.6$} & {\footnotesize{}$23.4$} &  & {\footnotesize{}$0.53$} & {\footnotesize{}$0.31$} & {\footnotesize{}$0.13$}\tabularnewline
{\footnotesize{}10} &  & {\footnotesize{}$58.7$} & {\footnotesize{}$42.3$} & {\footnotesize{}$23.6$} &  & {\footnotesize{}$0.53$} & {\footnotesize{}$0.31$} & {\footnotesize{}$0.13$}\tabularnewline
{\footnotesize{}11} &  & {\footnotesize{}$58.2$} & {\footnotesize{}$42.0$} & {\footnotesize{}$23.8$} &  & {\footnotesize{}$0.53$} & {\footnotesize{}$0.31$} & {\footnotesize{}$0.13$}\tabularnewline
{\footnotesize{}12} &  & {\footnotesize{}$57.8$} & {\footnotesize{}$41.8$} & {\footnotesize{}$24.0$} &  & {\footnotesize{}$0.53$} & {\footnotesize{}$0.31$} & {\footnotesize{}$0.13$}\tabularnewline
\hline 
\end{tabular}
\par\end{centering}{\footnotesize \par}
\end{singlespace}
\smallskip

\begin{singlespace}
{\footnotesize{}This table reports the sensitivity of the monthly
adjusted certainty equivalent return (in basis points) and the initial
stock allocation with respect to the investment horizon $N=2,3,...,12$
months, for a CRRA utility with $\gamma=5$ under different liquidity
settings $(\sigma_{\text{day}},\text{Vol}_{\text{day}})=\left(2.5,120m\right),(7.5,55m),(12.5,12m)$,
using $M=10^{5}$ Monte Carlo simulations with one control iteration
$I=1$. A portfolio of cash and SPDR S\&P 500 ETF is investigated,
with annual risk free rate $r^{f}=0.012$, portfolio weight increment
$0.01$ and initial investment amount $W_{0^{-}}=\$100m$.}{\footnotesize \par}
\end{singlespace}
\end{table}
\begin{table}[H]
{\small{}\caption{\label{tab:sen-gamma}Sensitivity to risk-aversion level}
}{\small \par}

\smallskip
\begin{singlespace}
\begin{centering}
{\footnotesize{}}%
\begin{tabular}{cclllllll}
 &  & \multicolumn{3}{l}{{\footnotesize{}CER}} &  & \multicolumn{3}{l}{{\footnotesize{}Initial allocation $\alpha_{0}$}}\tabularnewline
\hline 
\multirow{2}{*}{$\gamma$} & {\footnotesize{}$\sigma_{\text{day}}$} & {\footnotesize{}$2.5$} & {\footnotesize{}$7.5$} & {\footnotesize{}$12.5$} &  & {\footnotesize{}$2.5$} & {\footnotesize{}$7.5$} & {\footnotesize{}$12.5$}\tabularnewline
 & {\footnotesize{}$\text{Vol}_{\text{day}}$} & {\footnotesize{}$120m$} & {\footnotesize{}$55m$} & {\footnotesize{}$12m$} &  & {\footnotesize{}$120m$} & {\footnotesize{}$55m$} & {\footnotesize{}$12m$}\tabularnewline
\hline 
{\footnotesize{}2} &  & {\footnotesize{}$82.2$} & {\footnotesize{}$61.8$} & {\footnotesize{}$34.0$} &  & {\footnotesize{}$1.00$} & {\footnotesize{}$0.52$} & {\footnotesize{}$0.23$}\tabularnewline
{\footnotesize{}3} &  & {\footnotesize{}$72.6$} & {\footnotesize{}$53.4$} & {\footnotesize{}$29.5$} &  & {\footnotesize{}$0.81$} & {\footnotesize{}$0.42$} & {\footnotesize{}$0.19$}\tabularnewline
{\footnotesize{}4} &  & {\footnotesize{}$64.5$} & {\footnotesize{}$46.8$} & {\footnotesize{}$26.4$} &  & {\footnotesize{}$0.66$} & {\footnotesize{}$0.35$} & {\footnotesize{}$0.16$}\tabularnewline
{\footnotesize{}5} &  & {\footnotesize{}$57.8$} & {\footnotesize{}$41.8$} & {\footnotesize{}$24.0$} &  & {\footnotesize{}$0.53$} & {\footnotesize{}$0.31$} & {\footnotesize{}$0.13$}\tabularnewline
{\footnotesize{}6} &  & {\footnotesize{}$52.3$} & {\footnotesize{}$37.9$} & {\footnotesize{}$22.1$} &  & {\footnotesize{}$0.45$} & {\footnotesize{}$0.27$} & {\footnotesize{}$0.11$}\tabularnewline
{\footnotesize{}7} &  & {\footnotesize{}$47.6$} & {\footnotesize{}$34.7$} & {\footnotesize{}$20.4$} &  & {\footnotesize{}$0.38$} & {\footnotesize{}$0.23$} & {\footnotesize{}$0.10$}\tabularnewline
{\footnotesize{}8} &  & {\footnotesize{}$43.8$} & {\footnotesize{}$32.1$} & {\footnotesize{}$18.9$} &  & {\footnotesize{}$0.33$} & {\footnotesize{}$0.22$} & {\footnotesize{}$0.09$}\tabularnewline
{\footnotesize{}9} &  & {\footnotesize{}$40.5$} & {\footnotesize{}$30.0$} & {\footnotesize{}$17.8$} &  & {\footnotesize{}$0.29$} & {\footnotesize{}$0.20$} & {\footnotesize{}$0.08$}\tabularnewline
{\footnotesize{}10} &  & {\footnotesize{}$37.7$} & {\footnotesize{}$28.2$} & {\footnotesize{}$16.4$} &  & {\footnotesize{}$0.27$} & {\footnotesize{}$0.18$} & {\footnotesize{}$0.08$}\tabularnewline
\hline 
\end{tabular}
\par\end{centering}{\footnotesize \par}
\end{singlespace}
\smallskip

\begin{singlespace}
{\footnotesize{}This table reports the sensitivity of the monthly
adjusted certainty equivalent return (in basis points) and the initial
stock allocation with respect to the risk-aversion level $\gamma=2,3,...,10$
for a CRRA utility for investment horizon $N=12$ months, under different
liquidity settings $(\sigma_{\text{day}},\text{Vol}_{\text{day}})=\left(2.5,120m\right),(7.5,55m),(12.5,12m)$,
using $M=10^{5}$ Monte Carlo simulations with one control iteration
$I=1$. A portfolio of cash and SPDR S\&P 500 ETF is investigated,
with annual risk free rate $r^{f}=0.012$, portfolio weight increment
$0.01$ and initial investment amount $W_{0^{-}}=\$100m$.}{\footnotesize \par}
\end{singlespace}
\end{table}

\end{document}